\documentclass[traditabstract,longauth]{aa}
\usepackage{natbib}
\usepackage{epsfig}
\usepackage{graphicx}
\usepackage{amsmath}
\usepackage{amssymb}
\usepackage{hyperref}
\usepackage{natbib}
\usepackage{multirow}

\bibpunct{(}{)}{;}{a}{}{,}

\newcommand{\corot}{\textsl{CoRoT}}

\def\vsini{$v$\,sin\,$i$}             
\def\m2s2{\hbox{\,m$^{2}$\,s$^{-2}$}} 
\def\gcm3{\hbox{\,g\,cm$^{-3}$}}      
\def\vsini{\hbox{$v$\,sin\,$i_{\star}$}}      
\def\Msun{\hbox{$M_{\odot}$}}         
\def\Rsun{\hbox{$R_{\odot}$}}
\def\Mjup{\hbox{$\mathrm{M}_{\rm Jup}$}}
\def\Rjup{\hbox{$\mathrm{R}_{\rm Jup}$}}
\def\degr{\hbox{$^\circ$}}
\def\chisq{\mbox{$\chi^2$}}
\def\mp{{\emph M}$_{\rm p}$}
\def\rp{{\emph R}$_{\rm p}$}

\begin{document}

\title{Transiting exoplanets from the \corot~space mission XIV. \\
       CoRoT-11b: a transiting massive ``hot-Jupiter" in a\\ 
       prograde orbit around a rapidly rotating F-type star\thanks{The 
       \corot~space mission, launched on 2006 December 27, has 
       been developed and is operated by CNES, with the contribution 
       of Austria, Belgium, Brazil, ESA (RSSD and Science Programme), 
       Germany and Spain.}}


\author{  D.~Gandolfi\inst{\ref{Tautenburg},\ref{ESA}}
          \and G.~H\'ebrard\inst{\ref{IAP}}
          \and R.~Alonso\inst{\ref{Geneve}}
          \and M.~Deleuil\inst{\ref{LAM}}
          \and E.W.~Guenther\inst{\ref{Tautenburg}}
	  \and M.~Fridlund\inst{\ref{ESA}}
	  \and M.~Endl\inst{\ref{McD}}
	  \and P.~Eigm\"uller\inst{\ref{Tautenburg}}
          \and Sz.~Csizmadia\inst{\ref{DLR}}
          \and M.~Havel\inst{\ref{OCA}}
\and S.~Aigrain\inst{\ref{Oxford}} 
\and M.~Auvergne\inst{\ref{LESIA}} 
\and A.~Baglin\inst{\ref{LESIA}}
\and P.~Barge\inst{\ref{LAM}} 
\and A.~S.~Bonomo\inst{\ref{LAM}} 
\and P.~Bord\'e\inst{\ref{IAS}} 
\and F.~Bouchy\inst{\ref{IAP},\ref{OHP}} 
\and H.~Bruntt\inst{\ref{LESIA}}
\and J.~Cabrera\inst{\ref{DLR},\ref{LUTh}} 
\and S.~Carpano\inst{\ref{ESA}}
\and L.~Carone\inst{\ref{Koeln}}
\and W.~D.~Cochran \inst{\ref{McD}}
\and H.~J.~Deeg\inst{\ref{IAC}} 
\and R.~Dvorak\inst{\ref{Wien}} 
\and J.~Eisl\"offel\inst{\ref{Tautenburg}}
\and A.~Erikson\inst{\ref{DLR}}
\and S.~Ferraz-Mello\inst{\ref{Brasil}} 
\and J.-C.~Gazzano\inst{\ref{LAM},\ref{OCA}}
\and N.~P.~Gibson\inst{\ref{Oxford}}
\and M.~Gillon\inst{\ref{Geneve},\ref{Liege}} 
\and P.~Gondoin\inst{\ref{ESA}}
\and T.~Guillot\inst{\ref{OCA}}
\and M.~Hartmann\inst{\ref{Tautenburg}}
\and A.~Hatzes\inst{\ref{Tautenburg}} 
\and L.~Jorda\inst{\ref{LAM}} 
\and P. Kabath\inst{\ref{DLR},\ref{ESO-Chile}} 
\and A.~L\'eger\inst{\ref{IAS}} 
\and A.~Llebaria\inst{\ref{LAM}} 
\and H.~Lammer\inst{\ref{Graz}} 
\and P.~J.~MacQueen\inst{\ref{McD}}
\and M.~Mayor\inst{\ref{Geneve}}
\and T.~Mazeh\inst{\ref{Tel Aviv}} 
\and C.~Moutou\inst{\ref{LAM}} 
\and M.~Ollivier\inst{\ref{IAS}} 
\and M.~P\"atzold\inst{\ref{Koeln}} 
\and F.~Pepe\inst{\ref{Geneve}}
\and D.~Queloz\inst{\ref{Geneve}}
\and H.~Rauer\inst{\ref{DLR},\ref{ZAA}} 
\and D.~Rouan\inst{\ref{LESIA}}
\and B.~Samuel\inst{\ref{IAS}}
\and J.~Schneider\inst{\ref{LUTh}} 
\and B.~Stecklum\inst{\ref{Tautenburg}}
\and B.~Tingley\inst{\ref{IAC}} 
\and S.~Udry \inst{\ref{Geneve}}
\and G.~Wuchterl\inst{\ref{Tautenburg}}}

\institute{
     Th\"uringer Landessternwarte, Sternwarte 5, Tautenburg, D-07778 Tautenburg, Germany\label{Tautenburg}
\and Research and Scientific Support Department, ESTEC/ESA, PO Box 299, 2200 AG Noordwijk, The Netherlands\label{ESA} 
\and Institut d'Astrophysique de Paris, UMR7095 CNRS, Universit\'e Pierre \& Marie Curie, 98bis boulevard Arago, 75014 Paris, France\label{IAP}
\and Observatoire de l'Universit\'e de Gen\`eve, 51 chemin des Maillettes, 1290 Sauverny, Switzerland\label{Geneve}
\and Laboratoire d'Astrophysique de Marseille, 38 rue Fr\'ed\'eric Joliot-Curie, 13388 Marseille cedex 13, France\label{LAM}
\and McDonald Observatory, University of Texas at Austin, Austin, TX 78712, USA\label{McD} 
\and Institute of Planetary Research, German Aerospace Center, Rutherfordstrasse 2, 12489 Berlin, Germany\label{DLR}
\and Universit\'e de Nice-Sophia Antipolis, CNRS UMR 6202, Observatoire de la C\^ote d'Azur, BP 4229, 06304 Nice Cedex 4, France\label{OCA}
\and Oxford Astrophyiscs, Denys Wilkinson Building, Keble Road, Oxford OX1 3RH\label{Oxford}
\and LESIA, Observatoire de Paris, Place Jules Janssen, 92195 Meudon cedex, France\label{LESIA}
\and Institut d'Astrophysique Spatiale, Universit\'e Paris XI, F-91405 Orsay, France\label{IAS}
\and Observatoire de Haute Provence, 04670 Saint Michel l'Observatoire, France\label{OHP}
\and LUTH, Observatoire de Paris, CNRS, Universit\'e Paris Diderot; 5 place Jules Janssen, 92195 Meudon, France\label{LUTh}
\and Rheinisches Institut f\"ur Umweltforschung an der Universit\"at zu K\"oln, Aachener Strasse 209, 50931, Germany\label{Koeln}
\and Instituto de Astrof{\'i}sica de Canarias, E-38205 La Laguna, Tenerife, Spain\label{IAC}
\and University of Vienna, Institute of Astronomy, T\"urkenschanzstr. 17, A-1180 Vienna, Austria\label{Wien}
\and IAG, University of S\~ao Paulo, Brasil\label{Brasil}
\and University of Li\`ege, All\'ee du 6 ao\^ut 17, Sart Tilman, Li\`ege 1, Belgium\label{Liege}
\and European Southern Observatory, Alonso de Córdova 3107, Casilla 19001, Santiago de Chile, Chile\label{ESO-Chile}
\and Space Research Institute, Austrian Academy of Science, Schmiedlstr. 6, A-8042 Graz, Austria\label{Graz}
\and School of Physics and Astronomy, Raymond and Beverly Sackler Faculty of Exact Sciences, Tel Aviv University, Tel Aviv, Israel\label{Tel Aviv}
\and Center for Astronomy and Astrophysics, TU Berlin, Hardenbergstr. 36, 10623 Berlin, Germany\label{ZAA}
\and Dpto. de Astrof\'isica, Universidad de La Laguna, 38206 La Laguna, Tenerife, Spain\label{La Laguna}
\and Laboratoire d'Astronomie de Lille, Universit\'e de Lille 1, 1 impasse de l'Observatoire, 59000 Lille, France\label{Lille}
\and Institut de M\'ecanique C\'eleste et de Calcul des Eph\'em\'erides, UMR 8028 du CNRS, 77 avenue Denfert-Rochereau, 75014 Paris, France\label{IMCCE}
}
\date{Received ; accepted }

\abstract{The \corot~exoplanet science team announces the discovery of CoRoT-11b, a fairly massive 
          hot-Jupiter transiting a $V$=$12.9$~mag F6 dwarf star
	  (${\rm M_*}=1.27\pm0.05$~\Msun, ${\rm R_*}=1.37\pm0.03$~\Rsun, 
	  $T_\mathrm{eff}=6440\pm120$~K), with an orbital period of ${\rm P}=2.994329\pm0.000011$~days 
	  and semi-major axis ${\rm a}=0.0436\pm0.005$~AU. The detection of part of the radial velocity 
	  anomaly caused by the Rossiter-McLaughlin effect shows that the transit-like events 
	  detected by \corot~are caused by a planet-sized transiting object in a prograde orbit. The relatively 
	  high projected rotational velocity of the star ($\vsini=40\pm5$~km/s) places CoRoT-11 among the most 
	  rapidly rotating planet host stars discovered so far. With a planetary mass of \mp\,$=2.33\pm0.34$
	  \Mjup~and radius \rp\,$=1.43\pm0.03$~\Rjup, the resulting mean density of CoRoT-11b 
	  ($\rho_\mathrm{p}=0.99\pm0.15$~g/cm$^3$) can be explained with a model for an inflated hydrogen-planet 
	  with a solar composition and a high level of energy dissipation in its interior.
}

\titlerunning{The transiting exoplanet CoRoT-11b}
\authorrunning{Gandolfi et al.}

\maketitle

%
\section{Introduction}

Discovering and studying extrasolar planets, and in general planetary systems 
other than ours, aims at understanding whether the solar system is peculiar and 
unique or usual and unremarkable. In this context, the discovery of a large 
population of Jupiter-like planets with a semi-major axis $\lesssim0.1$~AU 
(i.e., hot-Jupiters), as well as the detection of massive planets in very 
eccentric orbits \citep{Moutou09a,OToole09} or even strongly misaligned with 
the stellar spin axis \citep{Hebrard08,Pont09}, has proven how the properties 
of extrasolar planets can be surprisingly different from those observed in 
the solar system's planets.

From this point of view, studies of transiting extrasolar planets are cornerstones 
for understanding the nature of planets beyond the solar system, since a wealth 
of precious information can be gained \citep{Winn10}. Indeed, the peculiar geometry 
of transiting planets makes them very special targets for obtaining direct 
measurements of the planet-to-star radius \citep[e.g.,][]{Rosenblatt71,Borucki84}. 
By combining time-series photometric observations acquired during the transit 
with radial velocity (RV) measurements of the host star, it is possible to derive 
the radius and mass of the planet, and therefore its mean density, once the mass 
and radius of the star has been determined. Transits also offer a unique 
opportunity to measure the sky-projected angle ($\lambda$) between the orbital 
angular momentum vector and the spin axis of the star \citep{Gaudi07}. This can 
be done by detecting the Rossiter-McLaughlin (RM) effect, i.e., the spectral 
distortion observed in the line profile of a rotating star as a second object 
passes in front of the stellar disc.

Furthermore transit surveys have the potential to enlarge the parameter space of 
planet host stars by detecting planets around stars that are usually not observed 
in radial velocity surveys. As is well known, Doppler surveys typically 
focus on slowly rotating solar-like stars because high RV precision can easily 
be achieved. They usually discard more massive main-sequence stars for which 
accurate RV measurements are rendered unfeasible by the rapid stellar rotation 
rate and the paucity of spectral lines. Indeed, a few RV searches have been conducted 
up to now around A- and F-type stars \citep[e.g.,][]{Lagrange09,Guenther09,Bowler10,
Johnson10}. Transit detections are not affected by the stellar rotation and can lead 
to the discovery of planets around rapidly rotating F-type stars. 

Space missions like \corot~\citep[Co\emph{nvection}, Ro\emph{tation, and planetary} 
T\emph{ransits},][]{Baglin06, Auvergne09} and \textsl{Kepler} \citep{Borucki10,Koch10} 
are crucial to increase the number of planets with well-known orbital and physical 
parameters, and consequently improve the database that is needed to investigate all 
the aspects of the exoplanets population, down to the Earth-like mass-regime. 
The recent discoveries announced by the \corot~exoplanet science team fully 
demonstrate the capability of the mission to determine the radius and the mean density 
of the transiting extrasolar planets, from the ``transition desert regime'' between 
brown dwarfs and planets \citep{Deleuil08}, across hot and temperate Jupiter-like 
objects \citep{Barge08,Alonso08,Bouchy08,Aigrain08,Moutou08,Rauer09,Fridlund10,Deeg10}, 
and down to the Earth-like radius regime \citep{Leger09,Queloz09}.

In the present paper the \emph{CoRoT Exoplanet Science Team} announces its eleventh 
transiting planet, namely \object{CoRoT-11b}, a fairly massive Jupiter-like planet in a 
relatively short-period orbit (about three days) around a rapidly rotating F6 dwarf star. 
By combining the high-precision photometric data from \corot~with RV measurements and high 
signal-to-noise spectroscopy from the ground, we fully characterised the planet's 
orbit and derived the main physical parameters of the planet-star system. Thanks to 
time-series RV measurements acquired during the transit we observed part of the RM 
effect and confirm the planetary transit event.

\section{\corot~observations, data reduction, and analysis}
\label{CoRoT-Obs}

\subsection{Satellite observations}
\label{Sec:Lightcurve}

CoRoT-11b was discovered during the \corot's second long run towards the 
Galactic centre direction, i.e., the \corot~\emph{LRc02} run.\footnote
{The `\emph{LR}' prefix means that the field is a long-run field (typically 
150 days of observations). The letter `\emph{c}' refers to the Galactic centre 
direction. The last two digits '\emph{02}' indicate that the observed field 
is the second \corot~long-run towards the Galaxy centre.} The observations 
lasted 145~days, from 2008 April 15 to September 7. The \emph{LRc02} field is 
centred at $\alpha\approx18^h\,42^m$ and $\delta\approx6\degr\,39\arcmin$ (J2000), 
between the \emph{Ophiuchus} and the \emph{Serpens Cauda} constellations. The 
planet was detected transiting the \corot~star\footnote{See \citet{Carpano09} for 
a full description of the \corot~target nomenclature.} with ID=0105833549. The main 
designations of the planet host star \object{CoRoT-11} along with its equatorial 
coordinates and optical and near-infrared photometry, are reported in Table~\ref{StarTable} 
as retrieved from the \emph{ExoDat} database \citep{Deleuil09} and \emph{2MASS} 
catalogue \citep{Cutri03}.

\begin{table}[!t]
  \caption{CoRoT, GSC2.3, USNO-A2, and 2MASS identifiers of the planet host star CoRoT-11. 
           Equatorial coordinates, optical, and near infrared photometry are from the \emph{ExoDat} 
	   catalogue \citep{Deleuil09} and \emph{2MASS} catalogue \citep{Cutri03}.}
  \centering
  \begin{tabular}{lll}       

  \multicolumn{1}{l}{Main identifiers}     \\
  \hline
  \hline
  \noalign{\smallskip}                
  \corot~ID        & 0105833549        \\
  GSC2.3 ID       & N1RO000587        \\
  USNO-A2 ID      & 0900-13499974     \\
  2MASS ID        & 18424494+0556156  \\
  \noalign{\smallskip}                
  \hline
  \noalign{\medskip}
  \noalign{\smallskip}                

  \multicolumn{2}{l}{Coordinates}     \\
  \hline                                  
  \hline                                  
  \noalign{\smallskip}                
  RA (J2000)      & $18^h\,42^m\,44\fs95$         \\
  Dec (J2000)     & $05\degr\,56\arcmin\,16\farcs12$  \\
  \noalign{\smallskip}                
  \hline
  \noalign{\medskip}
  \noalign{\smallskip}                

  \multicolumn{3}{l}{Magnitudes} \\
  \hline
  \hline
  \noalign{\smallskip}                
  \centering
  Filter & Mag & Error \\
  \noalign{\smallskip}                
  \hline
  \noalign{\smallskip}                
  B   & 13.596 & 0.024 \\
  V   & 12.939 & 0.019 \\
  $r^\prime$  & 12.638 & 0.019 \\
  $i^\prime$  & 12.283 & 0.053 \\
  J   & 11.589 & 0.021 \\
  H   & 11.416 & 0.029 \\
  Ks  & 11.248 & 0.022 \\
  \noalign{\smallskip}                
  \hline
  \end{tabular}
\label{StarTable}      
\end{table}

The transiting planet was detected after 51 days of observations in the 
so-called \corot~\emph{alarm-mode} \citep{Quentin06,Surace08}. 
This observing strategy consists in processing and analysing a first set of 
photometric data in order to single out planetary transits while the \corot~run 
is still on-going. This enabled us to switch the time sampling of the 
light curve from 512 to 32 seconds and trigger the ground-based follow-up 
observations (see Sect.~\ref{Follow-ups}). Thanks to an objective prism 
in the optical path of the \corot~exoplanet channel \citep{Auvergne09}, 
CoRoT-11 was observed in three broad-band colours (red, green, and blue), 
according to the specific photometric mask selected at the beginning of the 
run. This usually allows us to remove false positives that mimick planetary 
transit events, such as stellar activity or eclipsing binaries. A total of 261\,917 
photometric data-points were collected for each colour channel, 8\,349 of 
those were obtained with a time sampling of 512 sec, and 253\,568 with 
32 sec. The transit signal was detected in all three colour channels with 
the similar depth and the same duration and ephemeris as expected for a 
\emph{bona-fide} planetary transit.

\begin{figure*}[!t]
\begin{center}
 \epsfig{file=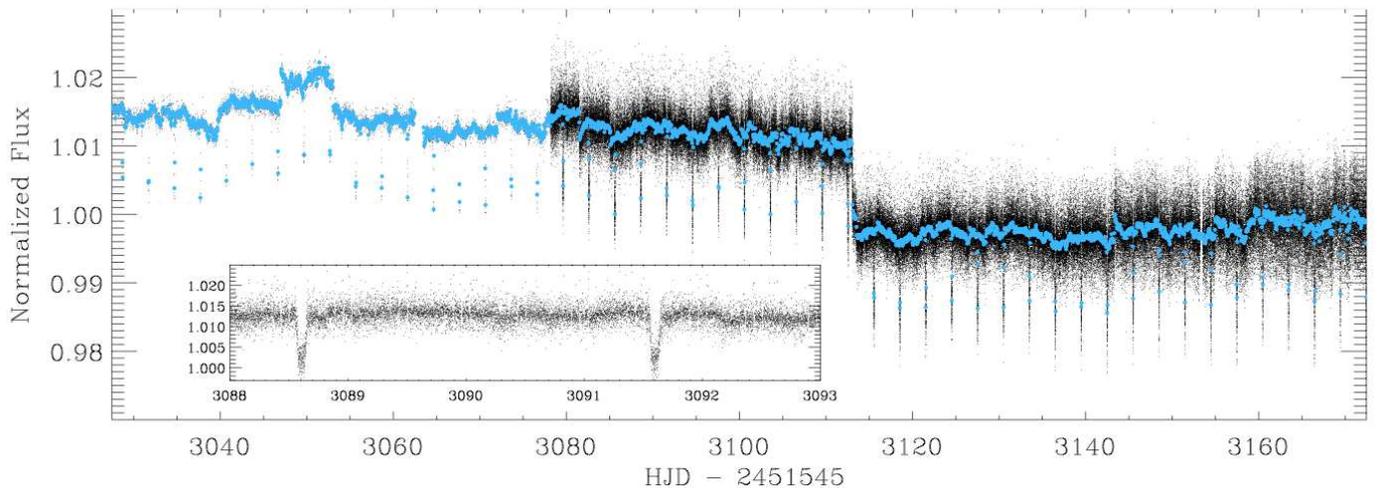,width=\textwidth, draft=false}
 \caption{Whole (145 days) cleaned light curve of CoRoT-11, sampled at 512 seconds 
          from the first $\sim50$ days, and at 32 seconds until the end of the run. 
	  For the sake of clarity, the blue dots represent 1-hour binned points. The 
	  inset plot is a zoom of the light curve showing two transits of CoRoT-11b. 
	  The ``jumps'' observed in the plot are caused by the impacts of high-energy 
	  particles onto the CCD's lattice. The light curve has been arbitrarily normalised 
	  to the median value of the flux. See the online edition of the Journal for a 
	  colour version of this figure.}
 \label{Lightcurve}
\end{center}
\end{figure*}

At the end of the \emph{LRc02} observing run, the whole photometric data-set of 
CoRoT-11 was processed using the \corot~reduction and calibration package described 
in \citet{Auvergne09}. The pipeline also flags bad photometric data-points collected 
during the entrance into and exit from the Earth's shadow or data-points that are 
strongly affected by hits of high-energy particles resulting from the crossing of the 
South Atlantic Anomaly (SAA).

\subsection{\corot~light curve and transit fit}
\label{Transit_Fit}

In order to increase the signal-to-noise (S/N) ratio, the analysis of the photometric 
data was performed using the white light curve, as derived by adding together the signal 
from the three colour channels. CoRoT-11 has a close neighbour star at about 
$2\arcsec$~northwest, falling well inside the \corot~photometric aperture, as well as a 
handful of faint nearby stars that are spatially located around the \corot~mask (see also 
Sect.~\ref{Phot-FU} and Fig.~\ref{CoRoT11b_imagete}). According to the $ExoDat$ database, 
the closest neighbour star is 2.1 magnitudes fainter than CoRoT-11 in the $r^\prime$-band. 
Following the method described in previous \corot~papers \citep[e.g.,][]{Alonso08}, we 
estimated that it contributes about $12.2$~\% of the total flux of the photometric mask of 
CoRoT-11. Taking also into account the light coming from the fainter neighbours, the total 
contamination fraction rises up to $13.0\pm1.5$~\%. This fraction was removed from the whole 
data-set prior to analysing the white light curve. 

A first cleaning of the raw \corot~data was performed applying an iterative sigma-clipping 
algorithm. Most of the photometric points rejected according to this criterion (about 7.8~\%) 
resulted from the crossing of the SAA and matched the photometric data previously flagged as 
outlier by the \corot~automatic pipeline \citep{Auvergne09}. The cleaned white light curve is 
plotted in Fig.~\ref{Lightcurve}. It shows 49 transits with a depth of $\sim1$~\%, occurring 
about every three days. The transits are clearly visible, as shown in the inset of 
Fig.~\ref{Lightcurve}. Out of 49 transits, 17 were detected with a 512 seconds time sampling, 
whereas 32 were observed with the oversampled rate (32 seconds). The light curve is moderately 
affected by sudden ``jumps'' in intensity resulting from the impact of high-energy particles 
onto the \corot~CCDs \citep{Pinheiro08}. The high-frequency scatter of the light curve 
(Fig.~\ref{Lightcurve}) is compatible with other sources of similar brightness observed by 
\corot~\citet{Aigrain09}.

The \corot~light curves are affected by a modulation of the satellite orbital period, which 
changes its shape and amplitude during a long run observation. Following the prescription of 
\citet{Alonso08}, the orbital signal of each $j^{th}$ orbit was corrected with the signals from 
the previous and the following 30 orbits. The data points acquired during the transits were not 
considered in the estimate of the mean orbital signal of the $j^{th}$ orbit. 

\begin{figure}[t]
\begin{center}
 \epsfig{file=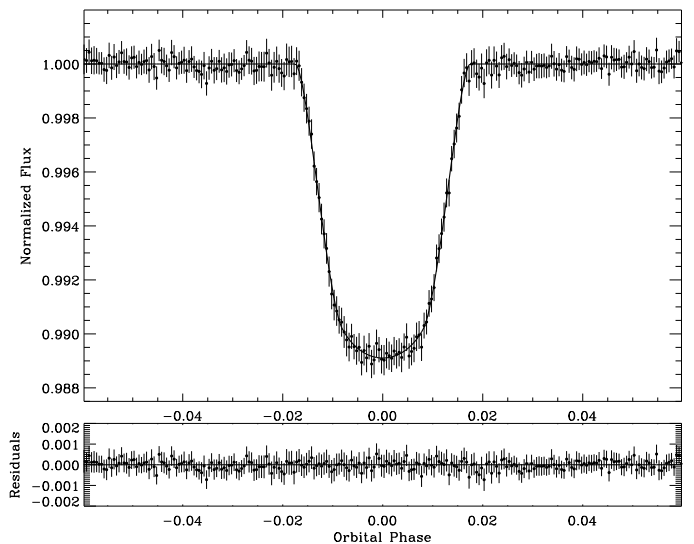,width=9cm}
 \caption{Binned and phase-folded curve of the transit of CoRoT-11b, with 
          the best-fit model over-plotted and the residuals of the fit. The 
	  standard deviation of the points outside transit is of 230 ppm 
	  (with a phase sampling of $5\times10^{-4}$, corresponding to about 
	  129~seconds). The mean error bar of the bins is of 239 ppm, revealing 
	  an insignificant amount of red noise in the phase-folded light 
	  curve after the corrections described in the text have been performed.
	  }
 \label{Transit}
\end{center}
\end{figure}

In order to determine the period $P$ and transit epoch $T_\mathrm{tr}$, we 
first used an approximate ephemeris to build a phase-folded curve of the 
transit. A simple trapezoidal model was fitted to this curve to get the 
parameters of the average transit, i.e., depth, duration, centre, and 
ingress/egress time. This model was then fitted to each individual transit, 
leaving only the centre as free parameter. A linear fit to the final 
observed-calculated ($O-C$) diagram of the transit centres served to refine the 
ephemeris, and we iterated the process until the fitted line had no 
significant slope. Once the orbital period and transit epoch were derived 
(Table~\ref{Par_table}), we constructed a combined phase-folded transit curve 
to this ephemeris by successively normalising each transit to the regions 
surrounding it. In this case, because the star is not very active (see 
Fig.~\ref{Lightcurve}), we used simple line fits to the phases between 
$[\,-0.06,-0.02\,]$, and $[\,0.02,0.06\,]$, using the ephemeris of 
Table~\ref{Par_table}, and applied this normalisation to the whole section from 
phases $[\,-0.06,0.06\,]$. We used a Savitzky-Golay filter to recognize a few 
remaining outliers before binning in phase. Taking into account the photometry 
rejected according to the first sigma-clipping algorithm, this process removed 
about $1.6$~\% of the remaining data points, leading to a final duty cycle of 
$\sim90$~\%.

Finally, the data-points were binned in blocks of 0.0005 in phase, and the error 
bars were estimated as the standard deviation of the data points inside each 
bin divided by the square root of the number of points inside the bin. The 
phase-folded curve of the transit is shown in Fig.~\ref{Transit}. 

The transit was fitted to a model using the formalism of \citet{Gimenez06}. To 
find the solution that best matches our data, we minimized the \chisq \, using 
the algorithm AMOEBA \citep{Press92}. The fitted parameters were the centre of 
the transit, the phase of start of the transit $\theta_1$, the planet-to-star 
radius ratio $k=R_\mathrm{p}/R_{*}$, the orbital inclination $i$ and the two 
non-linear limb darkening coefficients $u_{+}=u_a+u_b$ and $u_{-}=u_a-u_b$. 
We used a quadratic law for the limb darkening, given by 
$I(\mu)=I(1)[1-u_a(1-\mu)-u_b(1-\mu)^2]$, where I is the distribution of brightness 
over the star and $\mu$ is the cosine of the angle between the normal to the 
local stellar surface and the line of sight. The use of $u_{+}$ and $u_{-}$ is 
a better choice to avoid correlations between the two limb darkening coefficients 
$u_a$ and $u_b$, as described in \citet{Gimenez06}. To estimate the errors in 
each of the parameters we performed the \chisq minimization to five hundred 
different sets of data. These data-sets were constructed with different values 
for the contamination factor (estimated at $13.0\pm1.5$~\%), and different 
starting values to the AMOEBA minimization. To build each set, we first 
subtracted the best solution to the data, and then shifted circularly the residuals 
by a random quantity to keep the low-frequency content of the noise. The best-fit 
solution was finally added to these new residuals. For each of the fitted parameters 
in all data-sets, we calculated the standard deviation as an estimate of the errors. 
In order to take into account the effect of the non-Gaussian distributions of the 
parameters, we also forced a Gaussian fit to the measured distributions. The 
adopted (conservatively larger) error bar was the biggest error among the standard 
deviation of the fitted parameters and the $\sigma$ of the fitted Gaussian distribution.
The parameters and associated errors are listed in Table~\ref{Par_table},
along with the scaled semi-major axis $a/R_{*}$, as derived using Eq.~12 in
\citet{Gimenez06}. Assuming a circular orbit (i.e., $e=0$) and combining the scaled 
semi-major axis $a/R_{*}$ with the orbital period $P$ via the Kepler's laws, we 
derived the parameter $M^{1/3}_{*}/R_{*}=0.787\pm0.010$ in solar units, which leads 
to a mean stellar density $\rho_{*}=0.62\pm0.02$ g/cm$^3$ \citep[see][for the 
relevant formulas]{Seager03,Winn10}. The transit fit yielded  the limb darkening 
coefficients $u_+=0.61\pm0.06$ and $u_-=0.02\pm0.04$ (Table~\ref{Par_table}), in very 
good agreement with the theoretical values $u_+=0.64\pm0.01$ and $u_-=0.06\pm0.01$ 
predicted by \citet{Sing10} for a star with the same fundamental parameters as 
CoRoT-11 (Sect.~\ref{HR-spec}).

Note that the standard deviation of the residuals outside the transit phase is 230~ppm, 
which is within the uncertainties identical to the mean error bar of each of the bins 
(239~ppm), thus revealing the small low-frequency noise in the phase-folded light curve 
after the analysis described above.

\subsection{Planetary eclipse upper limits}

We searched for the eclipse of the planet in the \corot~light curve with the same 
techniques as described in \citet{Alonso09a, Alonso09b} and \citet{Fridlund10}. To 
account for a possible eccentric orbit, we mapped the \chisq\ levels of a fit to a 
trapezoid (with the shape parameters estimated from the transit) for different 
orbital phases. We did not obtain any significant detection above 100~ppm in depth.
We can thus set an upper 3\,$\sigma$ limit for the planetary eclipse depth of 100~ppm, 
which we translated into an upper limit to the brightness temperature of 2650~K 
\citep{Alonso09b}.

\section{Ground-based observations, data reduction, and analysis}
\label{Follow-ups}

As already described in previous \corot~discovery papers, 
intensive ground-based observations are mandatory to establish the planetary nature of the 
transiting candidates detected by \corot. These follow-ups are crucial to rule out possible false 
positives, i.e., physical configurations that mimick planetary transits, which cannot be excluded 
on the basis of meticulous light curve analyses only. Out of about $50$ promising candidates 
detected per \corot~long run field \citep[see][]{Carpano09,Moutou09b,Cabrera09,Carone10}, 
usually only a handful turns out to be \emph{bona fide} planetary objects. Furthermore, 
ground-based observations are needed to assess the planetary nature of the transiting object, 
derive the true mass of the planet, and measure the stellar parameters of the host star needed to 
accurately compute the planet radius\footnote{We remind the reader that transits provide the direct 
measurement of only the planet-to-star radius ratio.}.

Follow-up campaigns of the planetary transit candidates detected by the \emph{alarm-mode} in the 
\emph{LRc02} field started in early Summer 2008. In the following subsection we will describe 
the complementary photometric and spectroscopic ground-based observations of CoRoT-11.

\subsection{Photometric follow-ups}
\label{Phot-FU}

The objective prism placed along the optical path of the \corot~exoplanet channel spreads the light
of the observed targets over about 50 pixels on the CCDs, corresponding to a projected sky area of 
about $20\arcsec\times10\arcsec$. The total flux of each \corot~target is then computed by integrating 
the pixel signal over a preselected photometric mask elongated along the dispersion direction and 
covering most of the point spread function (PSF). As already described in Sect.~\ref{Transit_Fit}, 
this increases the possibility that the light of neighbour stars could contaminate the flux of the 
\corot~target. Furthermore, what is believed to be a ``good'' planetary transit might actually turn 
out to be the eclipse of a faint nearby binary system, whose light is diluted by the \corot~target star.

\begin{figure}[t] 
\begin{center}
\resizebox{\hsize}{!}{\includegraphics[angle=270]{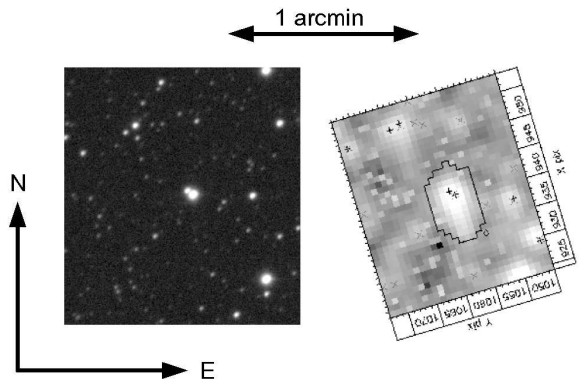}}
\caption{Sky area around CoRoT-11. \textit{Left}: $r^\prime$-filter image as retrieved from the $ExoDat$ database. 
                                                   CoRoT-11 is the brightest source in the centre of the image. 
						   The nearby contaminant star, about $2\arcsec$~northwest from 
						   the main target, is clearly visible. 
                                   \textit{Right}: image acquired by \corot, at the same scale and orientation. 
				                   The thick line around the target delimits the photometric mask 
						   used to integrate the signal of CoRoT-11. Note how the target 
						   and the contaminant are completely blurred and blended. The
	                                           crosses in the image mark the position of the stars in the field.}
\label{CoRoT11b_imagete}
\end{center}
\end{figure}

To reproduce the observed $\sim$$1.1$\,\% deep transit, a generic contaminant star cannot be more than 
$\sim$$5$~mag fainter than CoRoT-11. As already mentioned above, CoRoT-11 has a close, 2.1~mag fainter 
neighbour star that might thus actually be the {\it potential} source of false alarm (Fig.~\ref{CoRoT11b_imagete}). 
In order to exclude this scenario, we took advantage of the \corot~ephemeris to perform the so-called 
``on-off'' photometry. In this procedure, candidates are photometrically observed with ground-based 
facilities at higher spatial resolution than \corot during the transit (on-observation) and out of the 
transit (off-observation). The brightness of the candidates, as well as that of any nearby stars, is then 
monitored to unveil any potential background eclipsing binary. Full details of this method are described 
in \citet{Deeg09}.

\begin{figure}[t] 
\begin{center}
\resizebox{\hsize}{!}{\includegraphics{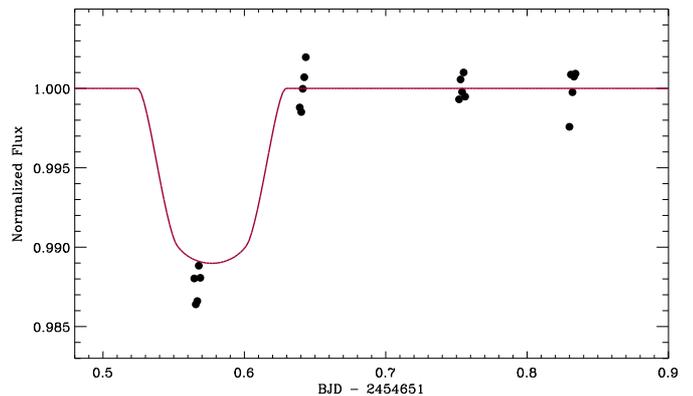}}
\caption{$r$-band light curve of CoRoT-11 as seen with the CCD camera mounted on the Swiss Leonard Euler 
         1.2\,m telescope at La Silla Observatory (Chile). The black circles mark each single exposure. The 
	 transit fit is over-plotted as derived from the \corot~light curve analysis.}
\label{EulerCam_LC}
\end{center}
\end{figure}

According to this observing strategy, $R$-band photometry was carried out on 2008 July 4, using the 
CCD camera mounted on the Swiss Leonard Euler 1.2\,m telescope at La Silla Observatory (Chile), 
under photometric sky condition. Four sets of five consecutive exposures of 45 seconds each 
were acquired. The first set was obtained during the predicted transit (on-observation), whereas the 
remaining three were taken out of transit (off-observations). The data were reduced with standard IRAF 
routines\footnote{IRAF is distributed by the National Optical Astronomy Observatory, which is operated 
by the Association of Universities for Research in Astronomy (AURA), inc., under cooperative agreement 
with the National Science Foundation.}; aperture photometry was performed with the DAOPHOT package 
under the IRAF environment. Differential photometry was obtained for CoRoT-11, as well as for the nearby 
contaminants, using a set of nearby comparison stars. The on-off Euler observations clearly show that 
CoRoT-11 is the source of the transit events detected by \corot~(Fig.~\ref{EulerCam_LC}). The first 
photometric data-set shows a dimming of the light of CoRoT-11 at the expected time and with roughly the 
same depth at mid-transit. The 2.1~mag fainter star located at about $2\arcsec$ northwest of the target 
should undergo eclipses with depths of about $0.1$~mag, something which the ground-photometry clearly 
excluded. By centring and co-adding the best-seeing Euler images, we excluded the presence of a third 
nearby object down to $R\approx17.5$~mag and up to $1.5\arcsec$ from CoRoT-11.

More Johnson $R$-band photometric observations of CoRoT-11 were carried out using the 30\,cm TEST telescope 
at the Th\"uringer Landessternwarte (TLS), Tautenburg (Germany), on 2008 September 7. Full details of the 
instrument, observing strategy, and data reduction can be found in \citet{Eisloffel07} and \citet{Eigmuller09}. 
Although these observations were performed at higher time-sampling than those at Euler, they were affected 
by poor weather conditions, especially in the second half of the night. Nevertheless, we succeeded to observe 
the transit ingress of CoRoT-11b at the expected \corot~ephemeris and exclude significant photometric variations 
in the nearby contaminant stars.   

\subsection{Reconnaissance low-resolution spectroscopy}
\label{LowRes_FU}

\begin{figure}[t] 
\begin{center}
\resizebox{\hsize}{!}{\includegraphics{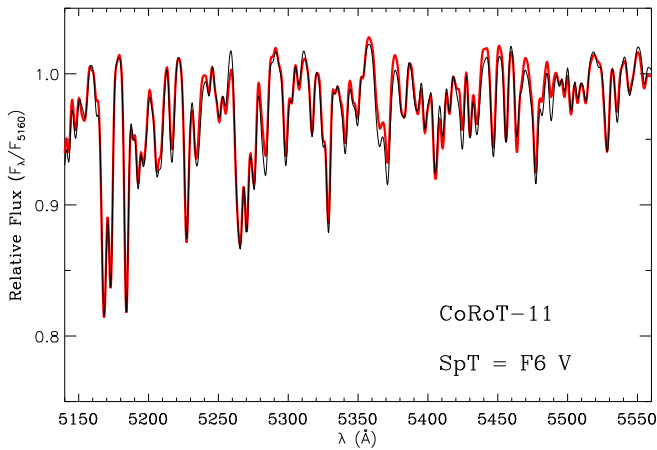}}
\caption{Section of the TLS low-resolution spectrum of CoRoT-11 (thin black line). Overplotted with a thick red 
         line is the best-fitting F6 V template. The spectra have been arbitrarily normalised to the flux at 
	 5160~\AA. See the online edition of the Journal for a colour version of this figure.}
\label{Fig_TLS}
\end{center}
\end{figure}

Low-resolution reconnaissance spectroscopy of the planet host star was performed with the long-slit spectrograph 
mounted at the Nasmyth focus of the 2\,m Alfred Jensch telescope of the TLS observatory, Tautenburg, Germany. These 
observations were part of an intensive programme devoted to the spectroscopic ``snap-shot'' of the planetary 
candidates detected by the \emph{alarm-mode} in the \emph{LRc02} \corot~field. They were useful to quickly classify 
the stars and derive a first estimate of their photospheric parameters. Furthermore, they allowed us to identify 
and remove giant stars, for which the transiting object would result in a low-mass stellar companion, as well 
as B-type objects and rapidly rotating early-type stars, for which high-precision radial velocity measurements 
cannot be achieved.

CoRoT-11 was observed on 2008 August 8, under clear and stable weather conditions. Three consecutive exposures 
of 20 minutes each were acquired and subsequently combined to remove cosmic ray hits and improve the 
S/N ratio. The data reduction was performed with a semi-automatic pipeline developed under the IDL\footnote{IDL 
is distributed by ITT Visual Information Solutions, Boulder, Colorado.} environment. 
Relative flux calibration was performed observing two spectro-photometric standard stars. The final extracted 
and co-added spectrum covers the wavelength range 4950--7320~\AA, with a resolving power $R\approx2100$ and 
an average $S/N\approx60$. The spectral type and the luminosity class of CoRoT-11 was derived by
fitting the observed spectrum with a grid of suitable template spectra, as described in \citet{Frasca03} and
\citet{Gandolfi08} and shown in Fig.~\ref{Fig_TLS}. We found that CoRoT-11 is an F6\,V star, with an accuracy
of about $\pm\,1$ sub-class (Table~\ref{Par_table}).

\subsection{Radial velocity observations}
\label{RV_FU}

The RV follow-up of the host star CoRoT-11 was started in summer 2008 by acquiring two high-resolution spectra 
with the SOPHIE spectrograph \citep{Bouchy09} attached to the 1.93\,m telescope of the Haute-Provence Observatory 
(France). The instrument was set in its high efficiency (HE), leading to a resolving power of $R\approx40\,000$. 
These observations revealed a rapidly rotating star with relatively broad spectral lines, corresponding to a 
projected rotational velocity (\vsini) of $\sim40$~km/s. According to the \corot~ephemeris, the SOPHIE spectra 
were secured around the extreme orbital phases (i.e., phase 0.25 and 0.75), and showed a RV variation of 
$\sim450$~m/s in phase with the \corot~ephemeris. Because the F6\,V spectral type of the host star translates into 
a stellar mass of about 1.3~\Msun, the measured RV variation is compatible with a $\sim2$~\Mjup~Jupiter planet. 
Twelve additional RV measurements were obtained in summer 2008 and 2009 using the HARPS spectrograph \citep{Mayor03} 
mounted at the 3.6\,m ESO telescope on La Silla (Chile). The spectra were acquired at different orbital phases, 
under good weather conditions and without strong moonlight contamination. The data were acquired setting the 
spectrograph both in the EGGS and HARPS standard modes, yielding a resolving power of $R\approx70\,000$ and 
$R\approx115\,000$, respectively.

The extraction of both the SOPHIE and HARPS spectra was performed using the respective pipelines. Following the 
techniques described by \citet{Baranne96} and \citet{Pepe02}, the radial velocities were measured from a weighted 
cross-correlation of the spectra with a numerical mask. We used a standard G2 mask that includes more than 3500 
lines. Cross-correlations with F0 and K5 masks gave similar results. One SOPHIE spectrum and three HARPS spectra 
were corrected for small moonlight contamination following the method described in \citet{Pollacco08} and 
\citet{Hebrard08}, which uses a reference background sky spectrum obtained through a second fiber spatially located 
near the target. This led to a radial velocity correction of $700\pm80$~m/s and below $450\pm50$~m/s for the SOPHIE 
and HARPS spectra, respectively.

\begin{table*}[!th]
  \centering 
  \caption{Radial velocities of the planet host star CoRoT-11 obtained with the SOPHIE, HARPS (EGGS and
  standard HARPS mode), COUD\'E@TLS, and HIRES spectrographs. The systemic velocities for each instrument,
  as derived from the circular Keplerian fit to the data, are reported on the right of the listed 
  spectrographs. The footnote indicates the RV measurements that have been corrected for moonlight contamination.}
  \label{Table_rv}
\begin{tabular}{crrrrc}
\hline
\hline
\noalign{\smallskip}
     HJD        &       RV~~    &$\sigma_{RV}$~~&   Bisector       & Texp   &    S/N per resolution   \\
    (days)      &     (km/s)    &  (km/s)       &    (km/s)~       & (sec)  &    element at 5500~\AA  \\
\noalign{\smallskip}
\hline
\noalign{\smallskip}
\multicolumn{1}{c}{SOPHIE - HE Mode} & & &\multicolumn{3}{r}{$V_r=-0.920\pm0.160$}~km/s \\
\hline
\noalign{\smallskip}
~~2454643.60252\tablefootmark{a}	&	-1.164	&   0.218   & &	    1402    &	    42       \\
2454683.42036	&	-0.714	&   0.206   &	&    1607     &	    50       \\
\noalign{\smallskip}
\hline
\noalign{\smallskip}
\multicolumn{1}{c}{HARPS - EGGS Mode} & & &\multicolumn{3}{r}{$V_r=-1.229\pm0.041$}~km/s \\
\noalign{\smallskip}
\hline
\noalign{\smallskip}
2454731.52836	&	-0.932	&   0.067   &	-0.531  &   2700    &	  88	   \\
2454742.51115	&	-1.364	&   0.104   &	-0.149  &   2700    &	  54	   \\
2454745.51100	&	-1.370	&   0.087   &	-0.476  &   2700    &	  66	   \\
2454746.51455	&	-0.986	&   0.071   &	-0.583  &   2700    &	  81	   \\
2454747.51953	&	-1.509	&   0.115   &	-0.892  &   1800    &	  48	   \\
\noalign{\smallskip}
\hline
\noalign{\smallskip}
\multicolumn{1}{c}{HARPS - Standard Mode} & & &\multicolumn{3}{r}{$V_r=-1.336\pm0.044$}~km/s \\
\noalign{\smallskip}
\hline
\noalign{\smallskip}
~~2455023.66206\tablefootmark{a}       &  -1.663  &  0.149 & -0.286 & 3600 &   34   \\
~~2455024.63997\tablefootmark{a}       &  -1.405  &  0.111 & -0.705 & 3600 &   48   \\
~~2455045.69605\tablefootmark{a}       &  -1.013  &  0.114 & -0.747 & 3600 &   52   \\
2455064.60739	                       &  -1.166  &  0.083 & -0.408 & 3300 &   64   \\
2455067.51277	                       &  -1.085  &  0.092 & -0.204 & 3300 &   59   \\
2455068.49041	                       &  -1.547  &  0.130 &  0.472 & 3300 &   42   \\
2455069.51299	                       &  -1.384  &  0.110 & -0.497 & 3300 &   48   \\
\noalign{\smallskip}
\hline
\noalign{\smallskip}
\multicolumn{1}{c}{COUD\'E@TLS} & & &\multicolumn{3}{r}{$V_r=-0.640\pm0.130$}~km/s \\
\noalign{\smallskip}
\hline
\noalign{\smallskip}
2455035.46055    &       -0.951  &   0.174   & & $2\times1800$ &       25       \\
2455057.40390    &       -0.541  &   0.175   & & $2\times1800$ &       25       \\
\noalign{\smallskip}
\hline
\noalign{\smallskip}
\multicolumn{1}{c}{HIRES} & & &\multicolumn{3}{r}{$V_r=-0.207\pm0.040$}~km/s \\
\noalign{\smallskip}
\hline
\noalign{\smallskip}
2455012.80417   &       -0.033  &   0.050   &  &     1200    &       61       \\
2455013.06216   &        0.181  &   0.086   &  &     1200    &       71       \\
2455013.78975   &       -0.165  &   0.089   &  &      900    &       54       \\
2455013.80073   &       -0.083  &   0.090   &  &      900    &       56       \\
2455013.81181   &       -0.170  &   0.135   &  &      900    &       56       \\
2455013.82275   &        0.173  &   0.111   &  &      900    &       57       \\
2455013.83360   &       -0.012  &   0.142   &  &      900    &       57       \\
2455013.84473   &        0.099  &   0.073   &  &      900    &       56       \\
2455013.85572   &        0.037  &   0.119   &  &      900    &       56       \\
2455013.86677   &        0.274  &   0.108   &  &      900    &       55       \\
2455013.87785   &        0.240  &   0.078   &  &      900    &       55       \\
2455013.88881   &        0.017  &   0.100   &  &      900    &       56       \\
2455013.89978   &        0.089  &   0.080   &  &      900    &       56       \\
2455013.91097   &       -0.263  &   0.086   &  &      900    &       54       \\
2455014.05389   &       -0.386  &   0.119   &  &      900    &       58       \\
\noalign{\smallskip}
\hline
\end{tabular}
\tablefoot{~\\
  \tablefoottext{a}{Corrected for moonlight contamination.}\\
}

\end{table*}

Two complementary RV measurements were acquired using the echelle spectrograph mounted 
on the 2\,m Alfred Jensch telescope of the TLS observatory, Tautenburg 
(Germany), in July and August 2009. For each observing night, two consecutive exposures of 
30 minutes each were recorded to increase the S/N ratio and remove cosmic ray hits. The adopted
instrument set-up yielded a spectral resolution of about $R\approx30\,000$. The data were 
reduced using standard IRAF routines. The wavelength solution was improved acquiring ThAr spectra 
immediately before and after each stellar observation. After accounting for instrumental shifts 
with telluric lines, the radial velocities were measured cross-correlating the target spectra with 
a spectrum of the RV standard star \object{HR\,5777} observed with the same instrument set-up.

As part of NASA's key science programme in support of the \corot~mission, more RV measurements 
were obtained with the HIRES spectrograph \citep{Vogt94} mounted on the Keck~I 10\,m telescope, 
at the Keck Observatory (Mauna Kea, Hawai'i). With the aim of detecting the Rossiter-McLaughlin 
effect, 13 RV measurements were secured during the expected transit on 2009 July 1 (UT).
The observations were performed with the red cross-disperser and the $I_2$ absorption 
cell to correct for instrumental shifts of the spectrograph. The $0\farcs861$ wide slit 
together with the 14$\arcsec$ tall decker was employed to allow proper background subtraction, 
yielding a resolving power of $R\approx50\,000$. In order to adequately sample the RM anomaly, 
the exposure time was set to 900 seconds. Two extra spectra of 1200 seconds each were also acquired 
out of transit, on 2009 June 30 (UT). The spectra were reduced with IRAF standard routines.  
The HIRES RV measurements were derived with the iodine data modelling code ``Austral" \citep{Endl00}.

The final RV measurements of CoRoT-11 are reported in Table~\ref{Table_rv}, along with error bars, 
exposure times, and S/N ratio. In spite of the good RV stability of the spectrographs used in the present 
work, the relatively high \vsini~of CoRoT-11 strongly affected the RV precision of our measurements and led 
to an accuracy in the range $\sim100-200$~m/s, with a typical error bar of about $100$~m/s even for the 
HARPS and HIRES data. The five data-sets, i.e., the SOPHIE, HIRES, and TLS data, and the two HARPS modes, 
were simultaneously fitted with a Keplerian model, assuming a circular orbit. The HIRES RV measurements 
acquired during the transit were not used in the fit. Both period and transit central time were fixed 
according to the \corot~ephemeris. An RV shift was let free to vary in the fit between the five data sets.

\begin{figure*}[th] 
\begin{center}
\includegraphics[scale=0.93]{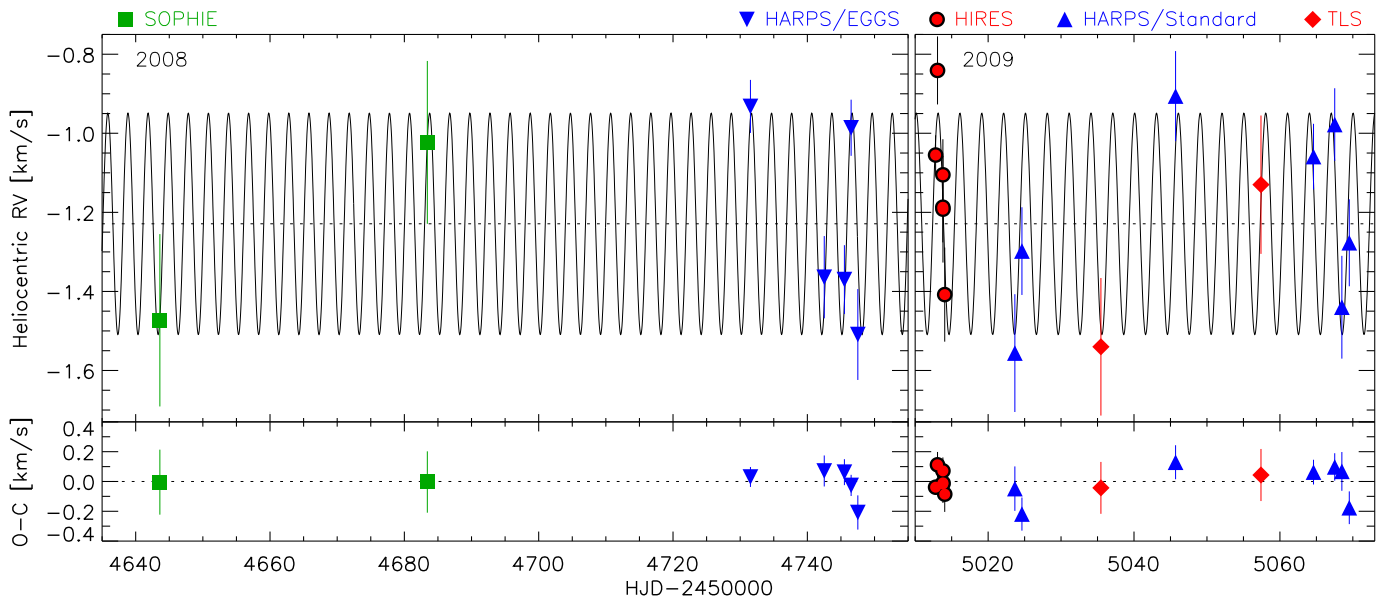}
\caption{\textit{Top:} Radial velocity measurements of CoRoT-11 with 1-$\sigma$\, error bars as a function of time and 
         the Keplerian fit to the data. The data are from SOPHIE (green squares), HARPS (blue downward and upward 
	 triangles for EGGS and HARPS modes, respectively), HIRES (red circles), and COUD\'E@TLS (red diamonds).
         The left and right panels show the two observational seasons in 2008 and 2009. The systemic radial velocity
	 of $V_r=-1.229\pm0.041$~km/s, as derived from the HARPS/EGGS data-set only, is plotted with a horizontal dotted 
	 line. \textit{Bottom:} Residuals of the fit.}
\label{Fig_omc}
\end{center}
\end{figure*}

The RV measurements are plotted in Fig.~\ref{Fig_omc} together with the best-fitting circular orbit. 
The derived orbital parameters are reported in Table~\ref{Par_table}, along with error bars that were 
computed from \chisq~variations and Monte~Carlo experiments. The RV measurements led to a semi-amplitude 
$K=280\pm40$~m/s. The standard deviation of the residuals to the fit is $\sigma_{O-C}=88$~m/s, in 
agreement with the expected accuracy of the RV measurements. The reduced \chisq~is 1.1 for the 22 RV 
measurements used in the fit. 

\begin{figure}[!th] 
\begin{center}
\includegraphics[scale=0.33]{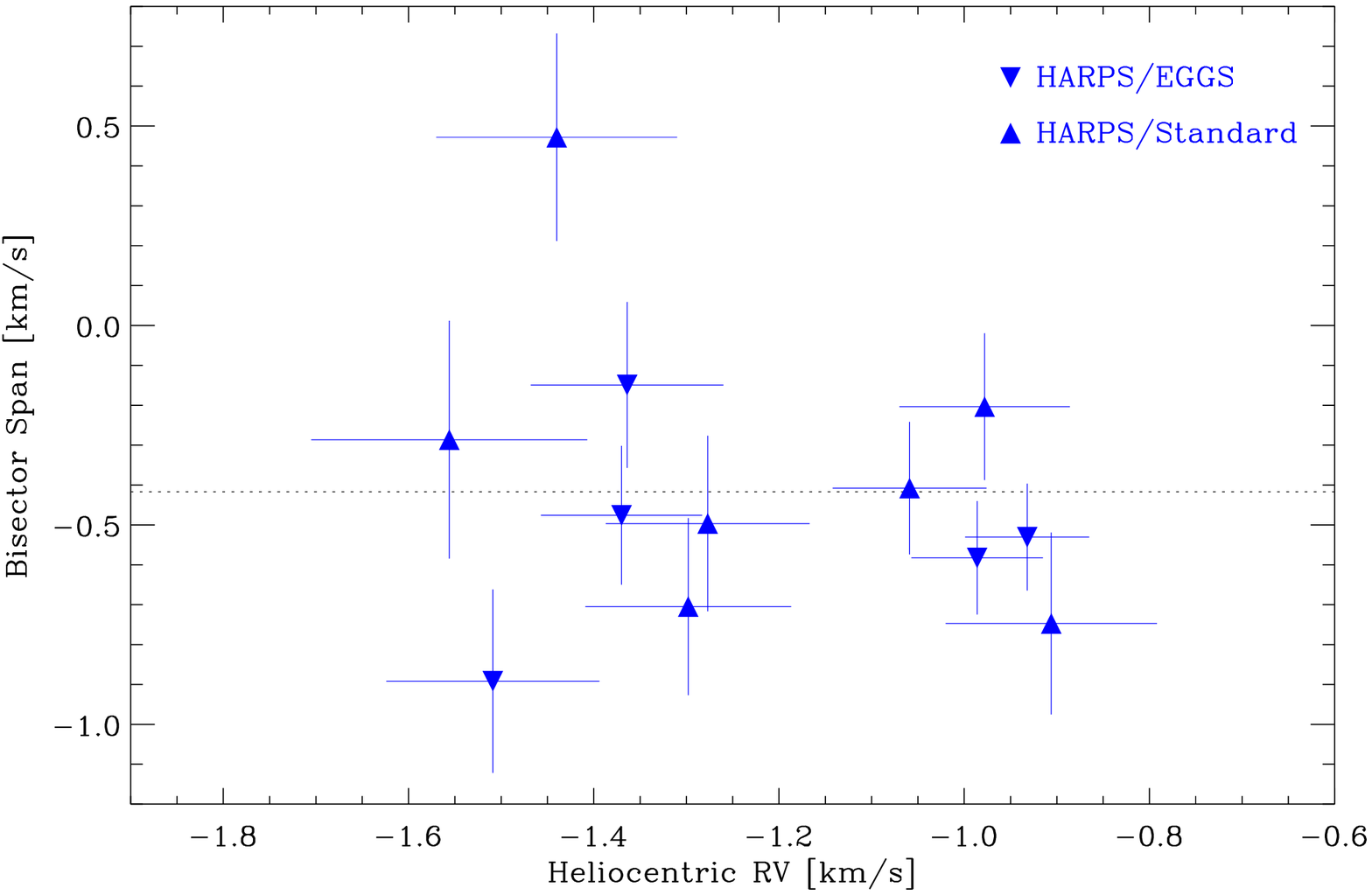}
\includegraphics[scale=0.33]{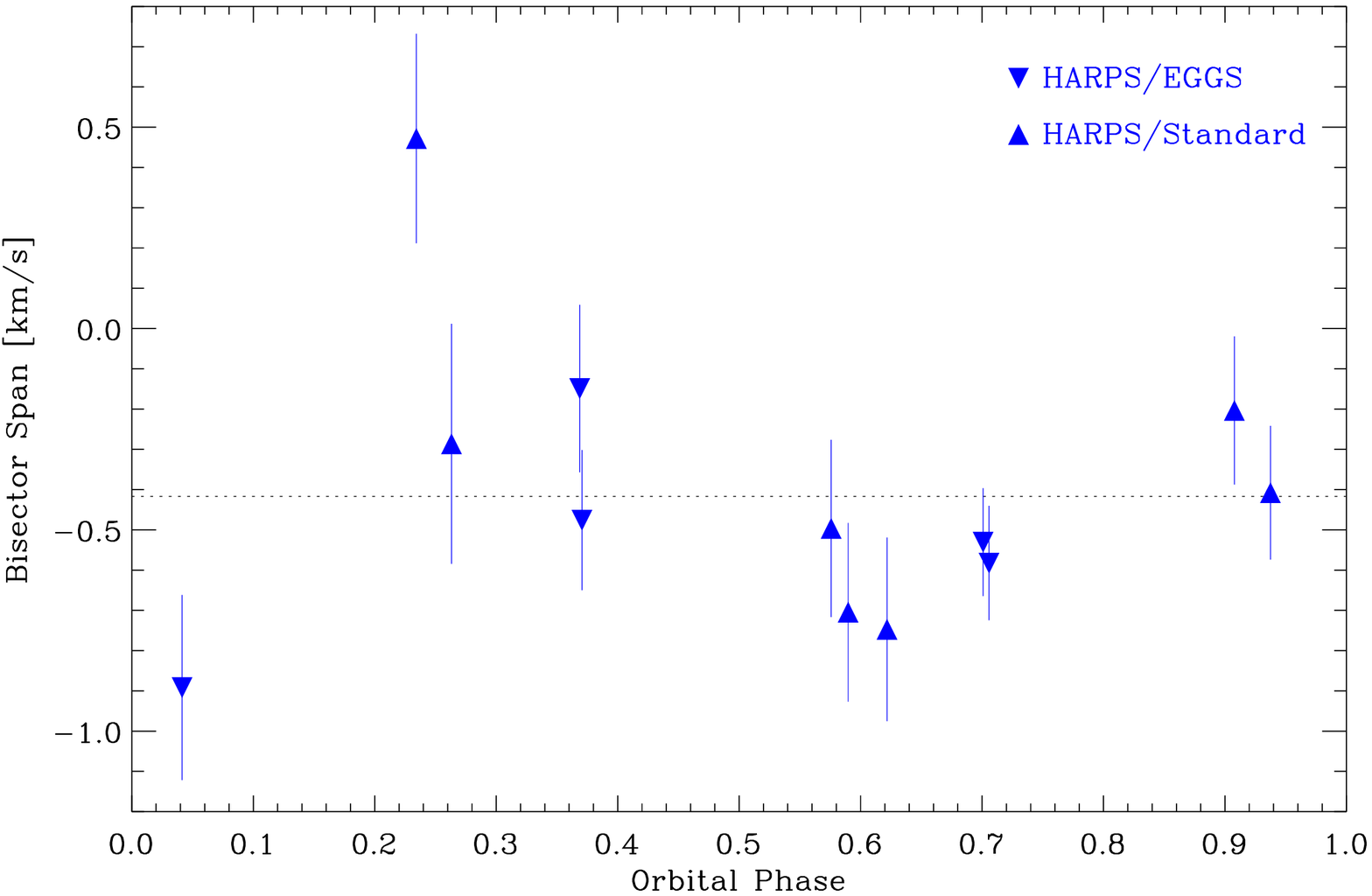}
\caption{Bisector spans versus radial velocity measurements (top panel) and orbital phases (bottom panel) as 
         derived from the HARPS data (blue downward and upward triangles for EGGS and HARPS modes, respectively). 
	 The horizontal dotted line marks the average negative value of the CCF bisector span, i.e. $-0.417$~km/s.}
\label{Bisector}
\end{center}
\end{figure}

We explored the possibility that the observed RV variations of CoRoT-11 do not result from the planet's 
orbital motion, but are instead caused by a periodic distortions in the spectral lines caused by either stellar 
magnetic activity or the presence of a hypothetical unresolved eclipsing binary, whose light is diluted by 
CoRoT-11. In order to exclude these scenarios, we performed an analysis of the cross-correlation function (CCF) 
profile. Using the highest resolution spectra in our data-set (i.e., the HARPS measurements) and following 
the line-bisector technique described in \citet{Queloz01}, we derived the difference in velocity space between 
the lower and upper part of the HARPS CCFs (i.e., bisector span). The value of the bisector span velocities are 
listed in Table~\ref{Table_rv}. The uncertainty was set to twice that of the corresponding HARPS radial 
velocity measurements. We found that the CCFs show a systematic asymmetric profile, translating into a negative 
value of the bisector span velocities (Table~\ref{Table_rv} and Fig.~\ref{Bisector}), which is usually observed in 
rapidly rotating F-type stars \citep{Gray86,Gray89}. Nevertheless, the CCF bisector spans show neither significant 
variations nor any trend as a function of both RV measurements and orbital phases (Fig.~\ref{Bisector}). The 
linear Pearson correlation coefficient between the HARPS RV measurements and the corresponding CCF bisector spans 
is $-0.25$. Removing the only outlier point with positive bisector span (i.e., $0.472$~km/s), the correlation 
coefficient approaches zero, being $-0.03$. Thus the observed RV variations seem not to be caused by spectral line profile 
variations to any significant degree, but are mainly due to the Doppler shift induced by the orbital motion of 
CoRoT-11b. The RV observations and the transit-signal detected by \corot~point to a hot-Jupiter-sized planet that 
orbits the star.

\begin{figure}[t] 
\begin{center}
\resizebox{\hsize}{!}{\includegraphics{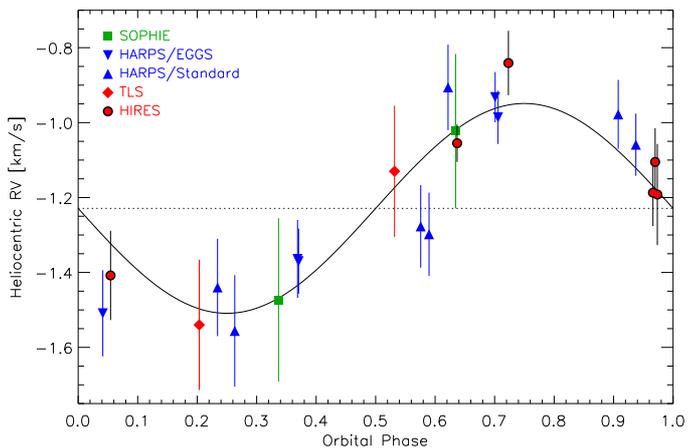}}
\caption{Phase-folded radial velocity measurements of CoRoT-11, and Keplerian fit to the data. The horizontal
         dotted line marks the systemic radial velocity $V_r=-1.229\pm0.041$~km/s, as derived from the 
	 HARPS/EGGS data-set only.  
}
\label{Fig_orb_phas}
\end{center}
\end{figure}

The phase-folded RV measurements are plotted in Fig.~\ref{Fig_orb_phas}. As already described, the orbit was 
assumed to be circular, which is a reasonable assumption for close-in hot-Jupiters. The radial velocities 
are not accurate enough to constrain the eccentricity with the orbital fit only. Indeed, a Keplerian fit with 
an eccentricity of about 0.6 provides a solution that agrees with the \corot~ephemeris, with a RV semi-amplitude 
$K$ which is 15~\% larger than the one obtained for a circular orbit. The standard deviation of the residuals 
to this eccentric fit ($\sigma_{O-C}=95$~m/s) is marginally higher than the circular fit. Only extremely eccentric 
orbits with $e>0.7$ produce low-quality fits, with dispersions larger than 120~m/s. Nevertheless, we put some
constraints on the possible value for the planet eccentricity, taking advantage of the transit fit parameter 
$M^{1/3}_{*}/R_{*}$, from which the mean stellar density can be inferred (Sect.~\ref{Transit_Fit}). The 
obtained value of $M^{1/3}_{*}/R_{*}$ depends on the eccentricity of the orbit. We found that for $e\gtrsim0.2$ 
the mean stellar density would be incompatible with a F6 dwarf star, because it is significantly higher than 
the expected value \citep{Cox00}.

\begin{figure}[!t] 
\begin{center}
\resizebox{\hsize}{!}{\includegraphics{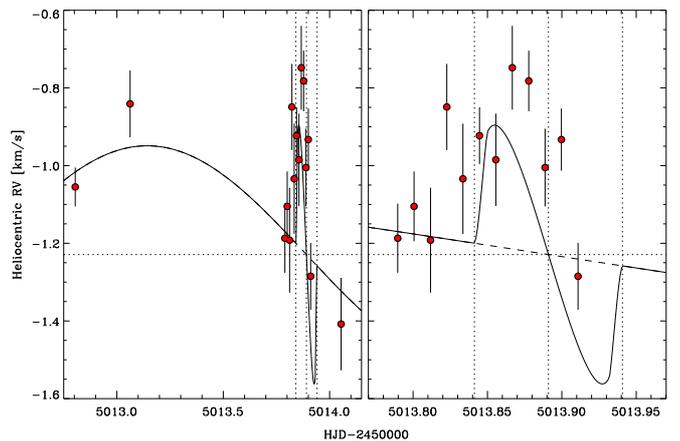}}
\caption{Radial velocities of CoRoT-11 measured with the HIRES spectrograph around 
the transit that occurred on July 1$^{st}$, 2010. The left panel shows 
all the RV data and the right panel shows a magnification on the transit. The 
dashed line shows the Keplerian fit without transit. The solid line shows the 
Rossiter-McLaughlin anomaly fit for $\lambda=0\degr$. The vertical dotted lines 
show the transit first contact, mid-time, and fourth contact. The data have been 
arbitrarily shifted to the systemic radial velocity of $V_r=-1.229\pm0.041$~km/s 
(horizontal dotted line), as derived from the HARPS/EGGS data-set only.
}
\label{Fig_RM}
\end{center}
\end{figure}

The on-transit RV data of CoRoT-11 are plotted in Fig.~\ref{Fig_RM}. Unfortunately, the HIRES 
observations were performed according to an old, slightly incorrect, transit ephemeris based on 
the analysis of the \emph{alarm-mode} data only. Nevertheless, although the HIRES measurements cover 
only the first half of the transit, they clearly show that the Rossiter-McLaughlin (RM) anomaly has 
been detected, which also confirms the occurrence of the transit events. The RM amplitude is large 
($\sim400$~m/s), because of the fast stellar rotation. This also proves that the transiting object has 
a planetary size. The first part of the spectroscopic transit shows radial velocities that are blue-shifted 
compared to the Keplerian fit, which clearly indicates that the orbit is prograde. In addition, 
systematics seem to be present in the data at a level above the expected uncertainties for some 
measurements. It is thus difficult to constrain the spin-orbit angle with the current data. In Fig.~\ref{Fig_RM} 
we show a model with $\lambda=0\degr$ and $\vsini=40$~km/s, using the analytical approach developed by 
\citet{Ohta05}. The fit is not satisfying, suggesting in particular a \vsini~value higher than the one 
we derived from the SOPHIE RV data and the spectral analysis (Sect.~\ref{HR-spec}), in order to have 
a larger amplitude for the anomaly. It is known that a discrepancy could be found between the 
\vsini~values measured from the RM effect and from the spectral modelling of line broadening, especially 
for fast rotators \citep[see e.g., ][]{Simpson10}. Concerning the spin-orbit angle, the data are 
compatible with $\lambda = 0\degr$. However, as for \vsini, the moderate quality of the data-set 
prohibits accurate measurements. Additional RM observations of CoRoT-11 should be performed, with a 
full coverage of the event.

\subsection{High-resolution spectroscopy and stellar analysis}
\label{HR-spec}

To derive the fundamental atmospheric parameters of the planet host star, 
we observed CoRoT-11 with the high-resolution spectrograph UVES mounted at 
the 8.2\,m Very Large Telescope (ESO-VLT; Paranal Observatory, Chile). Two 
consecutive spectra of 2380 seconds each were acquired in service mode on 2008 
September 17, under the ESO programme 081.C-0413(C). The star was observed through 
a $0\farcs5$ wide slit, setting the UVES spectrograph to its Dic-1 mode (346+580). 
The adopted configuration yielded a resolving power of about $65\,000$, with a 
spectral coverage $\lambda\approx3000-6800$~\AA. The spectra were extracted and 
combined with standard IRAF packages, giving a final S/N ratio of about 160 at 
5500~\AA. 

The effective temperature ($T_\mathrm{eff}$), surface gravity (log\,$g$), metallicity 
($[\rm{Fe/H}]$), and projected rotational velocity (\vsini) of CoRoT-11 were derived
following the procedure already adopted for other \corot~host stars \citep[e.g.,][]{
Deleuil08,Fridlund10,Bruntt10}. We took advantage of different spectral analysis packages 
applied independently by different teams within the \corot~community, e.g., the SME~2.1 
\citep{Valenti96,Valenti05}, the VWA \citep{Bruntt04,Bruntt08,Bruntt10} software. 
We found that the estimated values of the above mentioned physical 
parameters agree within the error bars. The final adopted values are $T_\mathrm{eff}=
6440\pm120$~K, log\,$g=4.22\pm0.23$, $[\rm{Fe/H}]=-0.03\pm0.08$, and $\vsini=40\pm5$~km/s 
(Table~\ref{Par_table}), with the latter value in perfect agreement with the one derived 
from the RV data (Sect.~\ref{RV_FU}).

\begin{figure}[t] 
\begin{center}
\resizebox{\hsize}{!}{\includegraphics[angle=90,scale=0.4]{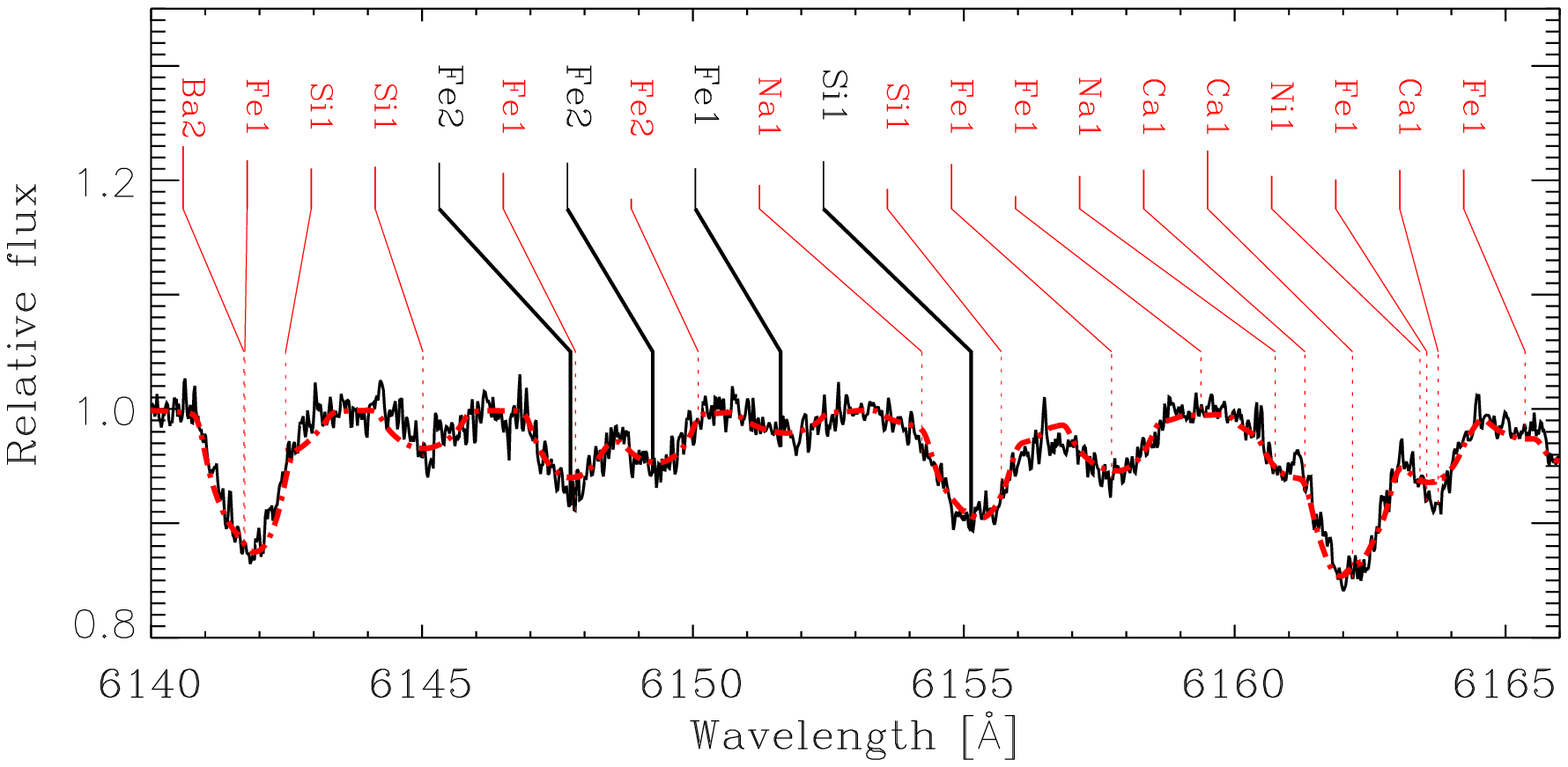}}
\caption{Small section of the observed UVES spectrum (thin black line) together
         with the best-fitting synthetic template (dashed line). The spectral lines
         used for the abundance analysis and the neighbouring lines are
         shown with black and red colours, respectively. See the online edition of 
	 the Journal for a colour version of this figure.}
\label{Fig_VWA}
\end{center}
\end{figure}

We also used the VWA software package to perform a detailed abundance analysis of the UVES
spectrum of CoRoT-11, by iteratively fitting reasonably isolated spectral lines. Atmosphere 
models were interpolated in a grid of MARCS models \citep{Gustafsson08} and atomic data 
were extracted from VALD \citep{Kupka99}. However, owing to the relatively high \vsini~ 
of the star, only 71 lines turned out to be sufficiently isolated and thus suitable 
for spectral analysis. A small section of the observed and fitted synthetic spectra is 
shown in Fig.~\ref{Fig_VWA}. Abundances were computed relative to the Sun to correct 
the oscillator strengths \citep[see][]{Bruntt08,Bruntt09}. We determined the atmospheric 
parameters by adjusting them to minimise the correlation of iron (Fe) with equivalent 
width (EW) and excitation potential (EP). Furthermore, we required that Fe\,{\sc i} 
and Fe\,{\sc ii} have the same mean abundance within the uncertainty. To evaluate 
the uncertainty on the atmospheric parameters, we perturbed them to determine when 
the correlations of Fe\,{\sc i} with EW or EP become significant or the Fe\,{\sc i} 
and Fe\,{\sc ii} abundances deviate by more than  $1\,\sigma$ \citep[see][for details]
{Bruntt08}. In Table~\ref{tab:ab} we list the abundances relative to the Sun
for the five elements Na, Si, Ca, Fe, and Ni.

\section{Results}
\label{Res}

\subsection{Stellar parameters of the parent star CoRoT-11}
\label{Star_Param}

To determine the mass and radius of the CoRoT-11 host star we took advantage of the stellar 
parameter ($M^{1/3}_{*}/R_{*}$) as derived from the \corot~light curve analysis 
(Sect.~\ref{Transit_Fit}), and of the effective temperature and metallicity 
($T_\mathrm{eff}$ and $[\rm{Fe/H}]$) as obtained from the spectral analysis 
(Sect.~\ref{HR-spec}). We thus compare the location of the star 
on a $log(M^{1/3}_{*}/R_{*})\,vs.\,log(T_\mathrm{eff})$ H-R diagram with evolutionary 
tracks computed with the CESAM code \citep{Morel08}. According to these theoretical 
models we obtained a stellar mass of $M_{\star}=1.27\pm0.05$~\Msun~and a stellar 
radius of $R_{\star}=1.37\pm0.03$~\Rsun, with an age of about $2.0\pm1.0$~Gyr 
(Table~\ref{Par_table}). From these results we 
derived a surface gravity of log\,$g=4.26\pm0.06$, in good agreement with the 
spectroscopically determined value of log\,$g=4.22\pm0.23$. We also checked whether 
the high rotation rate of the star can account for a significant flattening at the 
poles. According to the equation by \citet{Claret00} and assuming that the star is 
seen almost edge-on, the equatorial and polar radii should differ by only $\sim0.6$~\%, 
making the flattening effect negligible.

\begin{table}[t]
\centering
\caption{Abundances relative to the Sun for five elements in CoRoT-11. The number of 
         spectral lines used in the VWA abundance analysis are given in the last column.}
\label{tab:ab}
\begin{tabular}{lrr}
\hline
\hline
\noalign{\smallskip}
   Element     &       $[A/H]$~~~  &  $N$ \\
\noalign{\smallskip}
\hline
\noalign{\smallskip}
   {Na \sc  i} & $ -0.10\pm0.15  $ &   2  \\
   {Si \sc  i} & $  0.10\pm0.09  $ &   5  \\
   {Si \sc ii} & $  0.19\pm0.18  $ &   2  \\
   {Ca \sc  i} & $  0.03\pm0.12  $ &   4  \\
   {Fe \sc  i} & $ -0.04\pm0.08  $ &  44  \\
   {Fe \sc ii} & $ -0.02\pm0.08  $ &   6  \\
   {Ni \sc  i} & $ -0.08\pm0.11  $ &   8  \\
\noalign{\smallskip}
\hline
\end{tabular}
\end{table}

We computed the interstellar extinction to the star following the general guidelines 
described in \citet{Gandolfi08}. The seven $BVr^\prime i^\prime JHKs$ broad-band 
magnitudes as retrieved from the \emph{ExoDat} database enabled us to construct 
the spectral energy distribution (SED) of CoRoT-11, covering a wide spectral range, 
from optical to near-infrared wavelengths (see Table~\ref{StarTable}). Simultaneously 
using all the photospheric colours encompassed by the SED, we derived the 
interstellar extinction to the star ($A_{\mathrm V}$) by fitting the observed SED 
with a theoretical one progressively reddened with an increasing value of 
$A_{\mathrm V}$. The theoretical SEDs were computed with the \emph{NextGen} 
stellar atmosphere model \citep{Hauschildt99} with the same $T_\mathrm{eff}$, 
log\,$g$, and $[\rm{Fe/H}]$ as the star, the response curve of the $ExoDat$ 
photometric system, and the extinction law by \citet{Cardelli89}. Assuming a 
total-to-selective extinction $R_{\mathrm V}=A_{\mathrm V}/E_{\mathrm {B-V}}=3.1$ 
(typical of the diffuse interstellar medium in our Galaxy), as well as a black body 
emission at the star's effective temperature and radius, we derived an extinction 
$A_{\mathrm V}=0.55\pm0.10$~mag and a distance to the star $d=560\pm30$~pc 
(Table~\ref{Par_table}).

We attempted to derive the rotation period of the star from the \corot~light curve.
The Lomb-Scargle periodogram \citep{Scargle82} applied to the out-of transit 
data-points shows only one significant broad peak at about $8.5\pm2$~days, with a 
light curve amplitude of $\sim0.25$~\%. This period is not compatible with the maximum 
rotation period of $1.73\pm0.22$~days derived from the projected rotational velocity 
and radius of the planet host star. The detected signal at $8.5\pm2$~days might be 
due to a $0.02$~mag periodic variation of the nearby contaminant star located at about 
$2\arcsec$ from CoRoT-11 (see Sect.~\ref{Transit_Fit} and \ref{Phot-FU}). On the other 
hand, a low-level of magnetic activity of CoRoT-11 might account for no significant 
signals at $\lesssim2.0$~days, as confirmed from the absence of emission features in 
the core of the Ca\,{\sc ii} H \& K and Balmer lines.

\subsection{Planetary parameters of CoRoT-11b}
\label{Planet_Param}

Based on the stellar mass and radius (Sect.~\ref{Star_Param}), the RV curve 
semi-amplitude (Sect.~\ref{RV_FU}), the planet-to-star radius ratio, and the 
planet orbit inclination (Sect.~\ref{Transit_Fit}), we derived a mass 
for CoRoT-11b of \mp$=2.33\pm0.34$~\Mjup~and a radius of \rp$=1.43\pm0.03$~\Rjup, 
yielding a mean planetary density $\rho_\mathrm{p}=0.99\pm0.15$~g/cm$^3$. The planet 
orbits its host star at a distance of $a=0.0436\pm0.005$~AU in 
$2.994339\pm0.000011$~days. The planetary mass has been obtained assuming 
an eccentricity $e=0$. According to the results presented in Sect.~\ref{RV_FU}, we 
cannot exclude a slightly eccentric orbit, with $0\lesssim e \lesssim 0.2$. 
Nevertheless, for $e=0.2$ the planetary mass would decrease of about 4~\%, 
i.e., well within the error bar of our estimation. A summary of the planetary 
parameters derived in the present work is reported in Table~\ref{Par_table}.

%
\section{Discussion}
\label{Disc}

\begin{figure}[t] 
\begin{center}
\resizebox{\hsize}{!}{\includegraphics{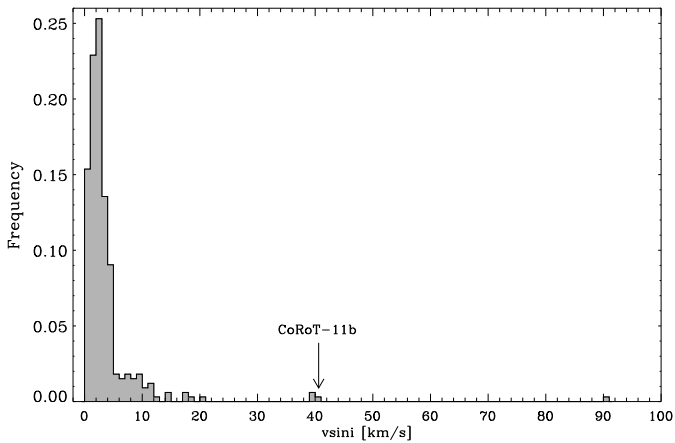}}
\caption{\vsini~distribution of the known extrasolar planets. The 
	 position of CoRoT-11b is highlighted with an arrow.}
\label{FIG_VsiniHist}
\end{center}
\end{figure}

Together with \object{30\,Ari\,Bb} \citep[][]{Guenther09}, \object{OGLE2-TR-L9b} \citep[][]{Snellen09}, 
and \object{WASP-33b} \citep[][]{CollierCameron10} orbiting a F6\,V ($\vsini=39$~km/s), 
F3\,V ($\vsini=39.3$~km/s), and A5\,V star ($\vsini=90\pm10$~km/s), respectively, 
CoRoT-11b is the fourth extrasolar planet discovered around a rapidly rotating main 
sequence star ($\vsini=40\pm5$~km/s; see Fig.~\ref{FIG_VsiniHist}). Furthermore, 
the planet host star CoRoT-11, with its effective temperature of 
$T_\mathrm{eff}=6440\pm120$~K, is one of the hottest stars known to harbour an 
extrasolar planet.

Most of the bulk of known extrasolar planets have been detected with 
the RV method. Although this technique has dramatically increased the number 
of discoveries in the last fifteen years, it suffers from a strong selection 
bias, because it is mostly restricted to planets around slowly rotating stars 
($\vsini\lesssim10$~km/s). This observational bias limits our knowledge of 
extrasolar planets to mainly late-type solar-like stars. One of the big 
advantages of the transit method is that it is insensitive to the stellar 
rotation, enabling us to single out planets even around intermediate-mass 
stars \citep{CollierCameron10}. This allows us to enlarge the 
parameter space of planet host stars and gives us a chance to study the planet 
formation around A and F stars. Even if CoRoT-11b has been confirmed and 
studied thanks to a complementary and intensive RV campaign, it would have 
likely been rejected from any RV search sample because of the fast rotation of its 
parent star. Indeed, about 20 RV measurements were needed to 
assess the planetary nature of CoRoT-11b and constrain its mass within 
$\sim15$~\%. But the RV signature of CoRoT-11b has been detected because 
of its high mass. Taking into account the accuracy of our RV measurements 
($100-200$~m/s), if the mass of CoRoT-11b had been \mp$\lesssim0.5$~\Mjup, 
it would not have been detected by the RV survey.    

\begin{figure}[t] 
\begin{center}
\resizebox{\hsize}{!}{\includegraphics{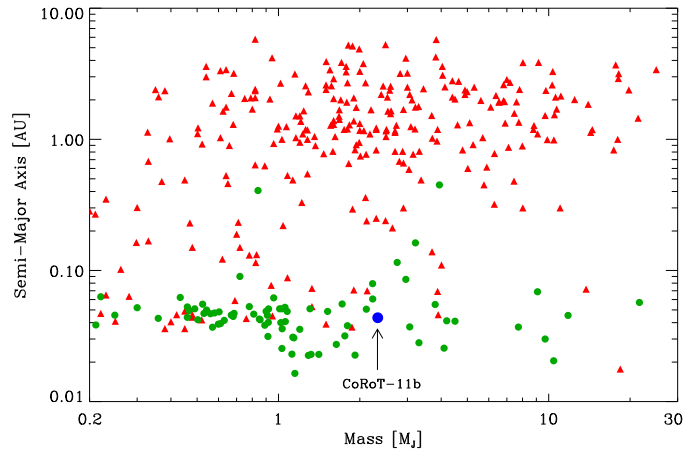}}
\caption{Semi-major axis versus planetary mass for the known extrasolar planets 
         detected in radial velocity (triangles) and transit surveys (circles). The 
	 position of CoRoT-11b is highlighted with an arrow.}
\label{FIG_AvsM}
\end{center}
\end{figure}

According to our planetary mass determinations (\mp$=2.33\pm0.34$), CoRoT-11b 
is among the most massive transiting hot-Jupiters discovered so far.      
It actually belongs to the poorly populated sub-group of objects with planetary masses 
around 2~\Mjup. As already noticed by \citet{Torres10}, there seems to be 
a lack of transiting hot-Jupiters with masses larger than about 2~\Mjup. Based on the 
list of currently known transiting planets\footnote{We refer the reader to the Extra 
Solar Planets Encyclopedia for a constantly updated list of known extra solar 
planets (http://exoplanet.eu/).}, hot-Jupiters with masses in the range 
$0.5\lesssim$\mp$\lesssim1.5$~\Mjup~seem to be $\sim5$ times more numerous than those 
with masses $1.5\lesssim$\mp$\lesssim2.5$~\Mjup. This trend is also confirmed by 
the number of planets discovered with the RV method. In Fig.~\ref{FIG_AvsM} the 
semi-major axis of the planets detected in radial velocity and transit surveys is 
plotted as a function of the planetary mass. Let us consider only the objects with 
$a\lesssim0.1$~AU and \mp$\gtrsim0.2$~\Mjup, i.e., hot-Saturn and Jupiter planets. 
There is a clear clump of hot-giant planets with masses between the mass of Saturn 
($\sim0.30$~\Mjup) and Jupiter, orbiting their parent star at about $0.04-0.06$~AU. 
Starting from $\sim1.0$~\Mjup, the number of hot-Jupiters seems to drop off, whereas 
the spread in the semi-major axis increases. For \mp$\gtrsim2$~\Mjup~the number of 
hot-Jupiters falls off significantly. The same trend is not seen for planets with  
$1\lesssim a \lesssim 5$~AU. The lack of Saturn planets orbiting their parent 
at such a distance might be owing to an observational bias of the RV technique. 
Nevertheless, since both the Doppler and transit methods are strongly sensitive in
detecting close-in massive planets, we conclude that hot-Jupiters with 
\mp$>2$~\Mjup~are significantly less common than ``normal'' hot-Saturn and Jupiter planets.

By assuming that CoRoT-11b is a hydrogen-rich gas giant we estimated the planet's thermal 
mass loss by applying the method outlined in detail in \citet{Lammer09}. Because the planet 
orbits a F6-type star with an age between 1.0 and 3.0 Gyr, we used the soft X-rays 
and EUV flux scaling law of Eq.~12 of \citet{Lammer09} and integrated the thermal mass 
loss during the planet's history up to the two age values given above. By using the stellar 
and planetary parameters and a heating efficiency $\eta$ for hydrogen-rich thermospheres, which 
can be considered between 10 - 25~\% \citep{Lammer09,Murray09}, we obtained a present time 
mass loss rate for CoRoT-11b of about $2.0\times10^{10}$ g/s or an integrated loss of 0.07~\% 
of its present mass ($\eta=10$~\%), and about $5.0\times10^{10}$~g/s, or 0.18~\% ($\eta=25$~\%) 
if the host star and planet are 1.0 Gyr old. If the the star/planet system is 3.0 Gyr old, we 
estimated a mass loss rate of about $3.0\times10^9$ g/s,  or $0.1$~\% ($\eta=10$~\%) and about 
$7.7\times10^9$~g/s, or 0.25~\% ($\eta=25$~\%) during the planet lifetime. These loss rates agree 
well with hydrodynamic escape model results for typical hot Jupiters \citep{Yelle08}. 
Although the planet radius is $1.43\pm0.03$~\Rjup, the main reason why the thermal mass loss of 
CoRoT-11b is not significant is related to the large mass of the planet of $2.33\pm0.34$~\Mjup. 
According to \citet{Lammer09}, only hot gas giant planets with $\rho_\mathrm{p}<<1$~g/cm$^3$ 
should experience large thermal mass loss.

\begin{figure}[t] 
\begin{center}
\resizebox{\hsize}{!}{\includegraphics{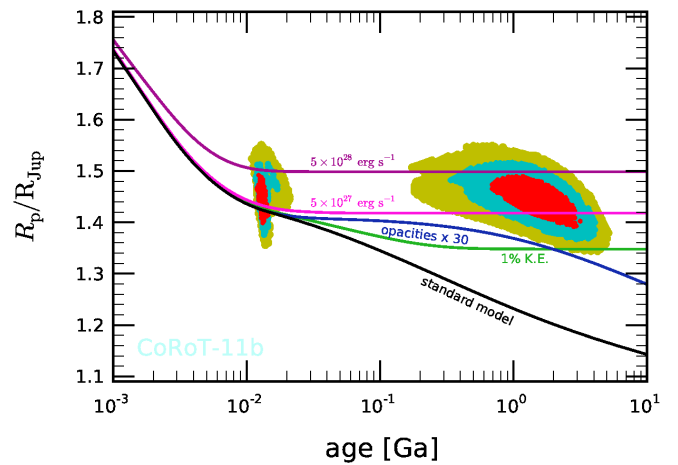}}
\caption{Evolution of the size of CoRoT-11b (in Jupiter units, $1\rm\,R_{Jup}=71\,492$~km) 
         as a function of the age of the system (in Ga=$10^9\,$years). The coloured areas  
         correspond to constraints derived from stellar evolutionary models matching the mean
	 stellar density and effective temperature within a certain number of standard 
	 deviations: less than $1\sigma$ (red),  $2\sigma$ (blue) or $3\sigma$ (green). 
	 The curves are evolutionary tracks for CoRoT-11b computed assuming a planetary mass 
	 of $M=2.33\rm\,M_{Jup}$ and equilibrium temperature $T_{\mathrm{eq}}=1657$~K), 
	 and using different models as labelled (see text for more details and the online 
	 edition of the Journal for a colour version of this figure).}
\label{Fig_Model}
\end{center}
\end{figure}

In order to investigate whether a standard model for an irradiated planet can account 
for the density of CoRoT-11b, we computed stellar and planetary evolution models using 
CESAM \citep{Morel08} and CEPAM \citep{Guillot95}, as described in \citet{Borde10} and 
\citet{Guillot10}. The results are shown in Fig.~\ref{Fig_Model} where the evolution 
of the size of CoRoT-11b is plotted as a function of the system age. The colours 
indicate the distance in standard deviations from the inferred effective temperature 
and mean stellar density, i.e., less than $1\sigma$ (red), $2\sigma$ (blue) or $3\sigma$ 
(green). These constraints are compared to planetary evolution models for a 
homogeneous solar-composition hydrogen-planet, with different hypotheses: (1) using 
a ``standard model'', i.e., without additional sources of heat; (2) by increasing interior 
opacities by a factor 30; (3) by adding a fraction ($\sim 1$\%) of the incoming 
stellar energy and dissipating it at the centre; (4) and (5) by dissipating 
$5\times 10^{27}$ and $5\times 10^{28}$~erg/s at the centre of the planet. 
The first three cases correspond to standard recipes used to explain the inflated 
giant exoplanets \citep{Guillot08}. The last two cases correspond to higher dissipation 
levels that are required to explain the planet size for the oldest ages.

Interestingly, as for CoRoT-2b \citep{Guillot10}, two  classes of solutions are found: 
(i) the standard solution for which the host star is on the main sequence (with an age 
of about $2.0\pm1.0$~Gyr) and the planet requires a high level of dissipation 
in its interior in order to account for its large size; (ii) a very young class of 
solutions in which the star is still on the pre-main sequence (PMS) phase (with an age of 
about $12\pm2$~Myr) and the planet size can be quite naturally explained  with a 
``standard model''. However, the latter scenario is in contradiction with the absence of 
the detectable Li\,{\sc i} $\lambda$6708~\AA~line in the spectra of CoRoT-11. According 
to our effective temperature determination ($T_\mathrm{eff}=6440\pm120$~K), CoRoT-11 
would belong to the narrow class of F-stars, which have suffered strong surface lithium 
depletion during the first billion years of their life. Indeed, studies of the lithium 
content in the photosphere of F-type stars in galactic clusters and field stars have 
revealed the presence of a narrow dip in the lithium abundance for effective temperature 
between 6500 and 6800~K \citep{Mallik03,Bohm04}. While the so-called ``lithium-dip'' 
is absent in the \object{Pleiades} \citep{Pilachowski87} and in general in all the young 
cluster ($\lesssim100$~Myr), this dip is well observed in older cluster like the 
\object{Hyades} \citep[$700$~Myr;][]{Boesgaard86a}, \object{NGC\,752} 
\citep[$1.7$~Gyr;][]{Hobbs86}, as well as in many field F-stars \citep[e.g.,][]{Boesgaard86b}. 
We thus believe that an age of $2.0\pm1.0$~Gyr is more plausible for the planet-star system. 

\begin{table*}
\centering
\caption{CoRoT-11b - Planet and star parameters.}            
\begin{tabular}{l r}
\hline
\hline
\noalign{\smallskip}
\multicolumn{2}{l}{\emph{Ephemeris}} \\
\noalign{\smallskip}
\hline
\noalign{\smallskip}
Planet orbital period $P$ [days]                         &      $2.994330\pm0.000011$   \\
Planetary transit epoch $T_\mathrm{tr}$ [HJD-2\,400\,000]  &  $54597.6790\pm0.0003$     \\
Planetary transit duration $d_\mathrm{tr}$ [h]             &      $2.5009\pm0.0144$     \\
\noalign{\bigskip}
\multicolumn{2}{l}{\emph{Results from radial velocity observations}}                  \\
\noalign{\smallskip}
\hline    
\noalign{\smallskip}
Orbital eccentricity $e$                                 &        0 (fixed)           \\
Radial velocity semi-amplitude $K$ [m/s]                 & $  280.0\pm40.0  $         \\
O-C residuals [m/s]                                      &         88                 \\
\noalign{\bigskip}
\multicolumn{2}{l}{\emph{Fitted transit parameters}} \\
\noalign{\smallskip}
\hline
\noalign{\smallskip}
Planet-to-star radius ratio $k=R_\mathrm{p}/R_{*}$       & $          0.1070\pm0.0005          $ \\
Linear limb darkening coefficient $u_+$                  & $            0.61\pm0.06            $ \\
Linear limb darkening coefficient $u_-$                  & $            0.02\pm0.04            $ \\
Orbital phase of planetary transit ingress ($\theta_1$)  & $         -0.0174\pm0.0001          $ \\
\noalign{\bigskip}
\multicolumn{2}{l}{\emph{Deduced transit parameters}} \\
\noalign{\smallskip}
\hline
\noalign{\smallskip}
Scaled semi-major axis $a/R_{*}$                         & $  6.890\pm0.080 $  \\
$M^{1/3}_{*}/R_{*}$ [solar units]                        & $  0.787\pm0.010 $  \\
Mean stellar density $\rho_{*}$ [g/cm$^3$]               & $   0.69\pm0.02  $  \\
Inclination $i$ [deg]                                    & $ 83.170\pm0.150 $  \\
Impact parameter\tablefootmark{a} $b$                    & $  0.818\pm0.008 $  \\
\noalign{\bigskip}
\multicolumn{2}{l}{\emph{Spectroscopic parameters}} \\
\noalign{\smallskip}
\hline
\noalign{\smallskip}
Effective temperature $T_\mathrm{eff}$ [K]               & $   6440\pm120  $ \\
Surface gravity log\,$g$\tablefootmark{b} [dex]          & $   4.22\pm0.23 $ \\
Surface gravity log\,$g$\tablefootmark{c} [dex]          & $   4.26\pm0.06 $ \\
Metallicity $[\rm{Fe/H}]$ [dex]                          & $  -0.03\pm0.08 $ \\
Stellar rotational velocity {\vsini} [km/s]              & $   40.0\pm5.0  $ \\
Spectral type                                            &        F6\,V\tablefootmark{d}      \\
\noalign{\bigskip}
\multicolumn{2}{l}{\emph{Stellar physical parameters from combined analysis}} \\
\noalign{\smallskip}
\hline
\noalign{\smallskip}
Star mass $M_{\star}$ [\Msun]                                             & $   1.27\pm0.05 $ \\
Star radius $R_{\star}$ [\Rsun]                                           & $   1.37\pm0.03 $ \\
Age of the star $t$ [Gyr]                                                 & $   2.0\pm1.0   $ \\ 
Interstellar extinction $A_{\mathrm V}$ [mag]                             & $   0.55\pm0.10 $ \\
Distance of the system $d$ [pc]                                           & $    560\pm30   $ \\
\noalign{\bigskip}
\multicolumn{2}{l}{\emph{Planetary physical parameters from combined analysis}} \\
\noalign{\smallskip}
\hline
\noalign{\smallskip}
Planet mass $M_\mathrm{p}$ [M$_\mathrm{J}$ ]\tablefootmark{e}             & $   2.33\pm0.34 $ \\
Planet radius $R_\mathrm{p}$ [R$_\mathrm{J}$]\tablefootmark{e}            & $   1.43\pm0.03 $ \\
Planet density $\rho_\mathrm{p}$ [g/cm$^3$]                               & $   0.99\pm0.15 $ \\
Orbital semi-major axis $a$ [AU]                                          & $ 0.0436\pm0.005$ \\
Equilibrium temperature\tablefootmark{f} $T_\mathrm{eq}$ [K]              & $   1657\pm55   $ \\
\noalign{\smallskip}
\hline       
\end{tabular}
\tablefoot{~\\
  \tablefoottext{a}{The impact parameter is defined as $b=\frac{a \cdot \cos{i}}{R_{*}}$}\\
  \tablefoottext{b}{Derived from the spectroscopic analysis.}\\
  \tablefoottext{c}{Derived using the light curve parameter $M^{1/3}_{*}/R_{*}$ and the stellar mass 
                     as inferred from stellar evolutionary models.}\\
  \tablefoottext{d}{With an accuracy of $\pm\,1$ sub-class.}\\		     
  \tablefoottext{e}{Radius and mass of Jupiter taken as 71492 km and 1.8986$\times$10$^{30}$ g, respectively.}\\
  \tablefoottext{f}{Zero albedo equilibrium temperature for an isotropic planetary emission.}
}
\label{Par_table}  
\end{table*}

\begin{acknowledgements}

We thank the anonymous referee for his/her careful reading, useful comments, and 
suggestions, which helped to improve the manuscript.

This paper is based on observations carried out at the European Southern Observatory 
(ESO), La Silla and Paranal (Chile), under observing programs numbers 081.C-0388, 
081.C-0413, and 083.C-0186.
The authors are grateful to the staff at ESO La Silla and ESO Paranal Observatories 
for their support and contribution to the success of the HARPS and UVES observing runs. 

This paper is also based on observations performed with SOPHIE at the Observatoire 
de Haute-Provence, France, under observing program PNP.08A.MOUT.

Part of the data presented herein were obtained at the W.M. Keck Observatory from 
telescope time allocated to the National Aeronautics and Space Administration through 
the agency's scientific partnership with the California Institute of Technology and the 
University of California. The Observatory was made possible by the generous financial 
support of the W.M. Keck Foundation. The authors wish to recognize and acknowledge the 
very significant cultural role and reverence that the summit of Mauna Kea has always 
had within the indigenous Hawaiian community. We are most fortunate to have the opportunity 
to conduct observations from this mountain. 

The team at IAC acknowledges support by grant ESP2007-65480-C02-02 of the Spanish Ministerio 
de Ciencia e Innovaci´on. The German \corot~Team (TLS and the University of Cologne) 
acknowledges DLR grants 50OW0204, 50OW0603, and 50QP07011.

This research has made use of the SIMBAD database, operated at CDS, Strasbourg, France.

\end{acknowledgements}

%

\listofobjects

\newpage

\end{document}